\documentclass[11pt,prd,a4paper,nofootinbib,groupedaddress]{revtex4-1}
\usepackage[english]{babel}
\pdfoutput=1

\usepackage{amsmath}
\usepackage{amsfonts}
\usepackage{amssymb}
\usepackage{xcolor}

\usepackage{physics}

\usepackage{graphicx}
\usepackage{float}
\usepackage[left=2cm,right=2cm,top=2cm,bottom=2cm]{geometry}

\begin{document}

\title{Free energy of the self-interacting relativistic lattice Bose gas at finite density}

\author{Olmo Francesconi}
\affiliation{Univ. Grenoble Alpes, CNRS, LPMMC, 3800 Grenoble, France}
\affiliation{Physics Department, College of Science, Swansea University (Singleton Campus), Swansea SA2
8PP, UK}

\author{Markus Holzmann}
\affiliation{Univ. Grenoble Alpes, CNRS, LPMMC, 3800 Grenoble, France}

\author{Biagio Lucini}
\affiliation{Mathematics Department, Computational Foundry, College of Science, Swansea University (Bay
Campus), Swansea SA1 8EN, UK}

\author{Antonio Rago}
\affiliation{Centre for Mathematical Sciences, University of Plymouth, Plymouth, PL4 8AA, UK}


\begin{abstract}
~\\
The density of state approach has recently been proposed as a potential route to circumvent the sign problem in systems at finite density. In this study, using the Linear Logarithmic Relaxation (LLR) algorithm, we extract the generalised density of states, which is defined in terms of the imaginary part of the action, for the self-interacting relativistic lattice Bose gas at finite density. After discussing the implementation and testing the reliability of our approach, we focus on the determination of the free energy difference between the full system and its phase-quenched counterpart. Using a set of lattices ranging from $4^4$ to $16^4$ , we show that in the low density phase, this {\em overlap free energy} can be reliably extrapolated to the thermodynamic limit. The numerical precision we obtain with the LLR method allows us to determine with sufficient accuracy the expectation value of the phase factor, which is used in the calculation of the overlap free energy, down to values of ${\cal O}(10^{-480})$. When phase factor measurements are extended to the dense phase, a change of behaviour of the overlap free energy is clearly visible as the chemical potential crosses a critical value. Using fits inspired by the approximate validity of mean-field theory, which is confirmed by our simulations, we extract the critical chemical potential as the non-analyticity point in the overlap free energy, obtaining a value that is in agreement with other determinations. Implications of our findings and potential improvements of our methodology are also discussed.
\end{abstract}

\maketitle

\section{Introduction}
Monte Carlo simulations of the system regularised on an Euclidean spacetime lattice provide the most efficient method for extracting quantitative information from non-supersymmetric non-Abelian gauge theories at zero density. The associated general methodology consists in generating configurations according to the Boltzmann weight $W(S) = e^{-S}$, with $S$ the Euclidean action of the system, and then computing averages of observables over the generated sample. In order for the method to work, the action $S$ needs to be real. This guarantees that the Euclidean path integral be positive definite, which allows us to interpret it as the partition function of an equivalent statistical system. However, there are physically relevant cases in which the action is complex and the Euclidean path integral becomes non-positive-definite. This gives rise to large numerical cancellations that generate noise overcoming by several orders of magnitude the typical signal one would like to observe, resulting in the inapplicability of importance sampling Monte Carlo for extracting physical observables. This cancellation, known in the literature as {\em the sign problem} (see e.g.~\cite{Gattringer:2016kco} for a recent review) characterises, among others, finite density systems in Quantum Field Theory and strongly correlated electron systems in Condensed Matter Physics. 

Resolving the difficulties caused by the sign problem would enable us to make substantial progress for systems such as finite density QCD, which at the moment can not be reliably studied either numerically or analytically. While a general algorithm that solves the sign problem for any system has been shown not to exist~\cite{Troyer:2004ge}, it is possible to devise numerical approaches that make the problem tractable in specific cases.\footnote{It is worth noting that for some systems, it has been shown that the sign problem disappears when the theory is reformulated in appropriate dual variables. While this may be a more general fact, currently the solution of the sign problem by dualization is possible only on prototype models. A recent review of this approach can be found in~\cite{Gattringer:2016kco}.} Recent examples include the Complex Langevin approach~\cite{Aarts:2008wh}, thimble regularisation~\cite{Cristoforetti:2013wha} and the density of states route, which, following the original proposal of~\cite{Gocksch:1988iz}  and further refinements~\cite{Anagnostopoulos:2001yb,Fodor:2007vv}, has been recently revisited in~\cite{Langfeld:2014nta,Gattringer:2015lra}.  Key to the latter two studies is the introduction of a restricted sampling~\cite{Langfeld:2012ah} in terms of the independent variable used to to define the density of states. This allows us to determine the logarithm of the density of states with exponential error reduction, hence  enabling us to perform extremely accurate measurements. If the precision of the determination of the density of states is high enough, one may eventually overcome cancellations that arise when computing numerical integrals. Examples of successful applications of the density of state method along these lines to systems affected by the sign problem have been provided in~\cite{Langfeld:2014nta,Gattringer:2015lra}.

Among theories used to test techniques to tame the sign problem, the self-interacting Bose gas at non-zero chemical potential is amongst the most widely studied. Here, we shall investigate this system at finite density as a function of the chemical potential using the density of state approach. Recently, this system has been studies with Complex Langevin~\cite{Aarts:2008wh} and dualization approaches~\cite{Endres:2006xu,Gattringer:2012df}, which, together with the good agreement with mean-field theory~\cite{Aarts:2009hn}, can be used to validate our results and hence to assess the viability of our proposal for the self-interacting Bose gas. These studies provide a scan in the parameters' space of the system and extract the associated phase structure. Using those results, we will fix the other action parameters to a set of values such that system is known to undergo a phase transition from zero particle net content to a dense phase for a critical value of the chemical potential. The density of states will be determined using the Linear Logarithmic Relaxation (LLR) algorithm~\cite{Langfeld:2012ah,Langfeld:2015fua}, which has been shown to provide exponential error reduction in a range of Lattice Gauge Theories applications involving e.g., tunnelling suppression at a first order phase transition~\cite{Langfeld:2015fua}, the determination of the free energy for a system with shifted boundary conditions~\cite{Pellegrini:2017iuy} and the de-correlation of the topological charge near the continuum limit~\cite{Cossu:2017sfu}. Earlier studies of complex action systems with the LLR and the closely related FFA method can be found respectively in~\cite{Langfeld:2014nta,Lucini:2014wga,Langfeld:2016mct,Langfeld:2016kty,Garron:2016noc,Garron:2017fta}~and~\cite{Mercado:2014dva,Gattringer:2015lra,Giuliani:2016tlu,Giuliani:2017fss} (see also~\cite{Bloch:2018yhu}). In this work, we will focus on the full - phase quenched overlap free energy. This quantity  is defined as the free energy difference between the original system and the system obtained by setting to zero the imaginary part of the action, the latter being referred to as {\em phase quenched system}. The reason for choosing this observable is twofold: (1) as it will be shown below, the overlap free energy controls the severity of the sign problem; (2) in the formulation of the density of states used in this work, this free energy is a central quantity to determine the density of particles (see e.g.~\cite{Langfeld:2014nta}). Given its characteristic exponential error reduction, the LLR algorithm provides an efficient method to compute this free energy, or equivalently the ratio of the two partition functions that defines this observable.  The purpose of this work is to understand whether the method we are introducing can be used for characterising the phases of the system and if the numerical measurements are precise enough for studying the phase transition that occurs at a critical value of the chemical potential. 

The rest of the paper is organised as follows. In Sect.~\ref{sect:dos} we review the density of state method and we discuss the main observables targeted in our study. Sect.~\ref{sect:bose} describes the self-interacting Bose gas at finite chemical potential on a spacetime lattice. A description of the numerical methodology used in our work is reported in Sects.~\ref{sect:llr} to~\ref{sect:bias_opt}. Numerical results are presented and discussed in Sect.~\ref{sect:res}. Finally, Sect.~\ref{sect:conclusions} contains a critical discussion of our findings and their implications. Some earlier results related to the current study have been reported in~\cite{Pellegrini:2015dkk}~and, more recently, in~\cite{Lucini:2019abc}.

\section{Generalised density of states} \label{sect:dos}
The generalised density of states method provides a straightforward approach to the problem of simulating systems with a complex action
\begin{equation}\label{eq:action}
S[\phi]=S^R[\phi]+i\mu S^I[\phi],
\end{equation}
where we have explicitly separated the real and imaginary part.
For convenience, we have assumed a linear dependence of the latter on the chemical potential, and
suppressed any parameter dependence of $S^R$ and $S^I$, which, in particular, may also depend on $\mu$. 
Without loss of generality, any partition function can be written as a functional
integral over the fields,
\begin{equation} \label{eq:part}
Z(\mu)=\int {\cal D}\phi \ e^{-S^R[\phi]-i\mu S^I[\phi]}.
\end{equation}
Introducing a generalised density of states (DoS) function,
\begin{equation}\label{eq:dos_def}
\rho(s) = \int {\cal D}\phi \ \delta(s-S^I[\phi]) \ e^{-S^R[\phi]},
\end{equation}
the  partition function can be obtained from a 1-dimensional integration
\begin{equation} \label{eq:part_dos}
Z(\mu)= \int \ \rho(s) \ e^{-i\mu s} \ \dd s.
\end{equation}
This reformulation suggests to split up the problem of evaluating the partition function of systems with complex action in two separate steps: first, to evaluate $\rho(s)$ numerically to a high level of precision, and then tackle the influence of the imaginary part of the action separately by performing the remaining one dimensional integral. Although we still expect a sign problem manifesting from the need of  cancellations over multiple orders of magnitude in the oscillatory integral, we have transformed a multidimensional oscillatory integration to a \textit{softer} variant where the resulting one dimensional Fourier transform is separated from the Monte Carlo integration.

The theory obtained by neglecting the imaginary part of the action is usually referred to as "phase quenched" and the associated partition function is given by
\begin{equation} \label{eq:part_pq}
Z_{pq} = \int {\cal D}\phi \ e^{-S^R(\phi)} = \int \rho(s) \ \dd s.
\end{equation}
It is worth noting that the phase-quenched system can be studied via standard importance sampling techniques, as it is a \textit{sign problem}-free system.
However, the physics described by it is not a good representation of the physics of the full system.

To evaluate the hardness of the sign problem it is possible to evaluate the overlap factor between the full and phase quenched theory defined as the ratio of the two partition functions
\begin{equation}
\frac{Z}{Z_{pq}}
=
\frac
{\int {\cal D}\phi \ e^{-S^R[\phi]} \ e^{-i\mu S^I[\phi]}}
{\int {\cal D}\phi \ e^{-S^R[\phi]}}.
\end{equation}
We can interpret this quantity as the expectation value of the phase in the phase quenched theory, from here on defined as $\langle e^{i \varphi} \rangle_{pq}$. 
Thanks to the symmetry $\rho(s)=\rho(-s)$ of the DoS, the phase factor is obtained from the real part of the Fourier transform of the DoS
\begin{equation} \label{eq:mv_phase}
\langle e^{i \varphi} \rangle_{pq} =
\frac{Z}{Z_{pq}} = 
\frac{\int \rho(s) \cos(\mu s) \dd s}{\int \rho(s) \dd s}.
\end{equation}

Physically, $\langle e^{i \varphi} \rangle_{pq}$ is related to the free energy difference between the full and phase quenched systems. Specifically, writing
\begin{equation}
Z = e^{ - F V} \qquad \mbox{and} \qquad Z_{pq} = e^{- F_{pq} V} , 
\end{equation}
with $F$ ($F_{pq}$) being the free energy per unit of volume $V$ of the full (phase quenched) system, we have that 
\begin{equation} \label{eq:df}
\Delta F = F - F_{pq} = -\frac{1}{V} \log{\langle e^{i \varphi} \rangle_{pq}}. 
\end{equation} 

Since $\langle e^{i \varphi} \rangle_{pq} \le 1$, the phase quenched model provides a lower bound
for the free energy.
To provide the expected finite $\Delta F$ in the thermodynamic limit,
$|\log{\langle e^{i \varphi} \rangle_{pq}}| \propto V$, hence $\langle e^{i \varphi} \rangle_{pq}$ has to be exponentially small in $V$, implying that the oscillatory integral \eqref{eq:mv_phase} that defines it must provide cancellations over many orders of magnitude. 

In our density of states approach for complex action systems, we will show that high precision data for the discretised DoS can be obtained by specialised Monte Carlo methods as provided by the linear logarithmic relaxation (LLR) algorithm. However, in order to obtain a precise and numerically stable evaluation of the Fourier integral,  the full continuous DoS must be reconstructed. The relativistic Bose gas studied in this work provides a concrete frame of application for our methodology and at the same time a probing benchmark to assess how effective it can be. 

\section{Model: Relativistic Bose Gas} \label{sect:bose}
In this paper we will concentrate on the relativistic Bose gas. This model has been extensively studied in the context of the sign problem of lattice field theories via Complex Langevin dynamics and mean-field approximation \cite{Aarts:2009hn} as well as complete dualization \cite{Gattringer:2012df}. Therefore, it is a good candidate for a benchmark study to test whether or not the DoS approach can be used to mitigate the sign problem in lattice field theories.

The lattice discretised action can be expressed as:
\begin{equation} \label{eq:bose_disc}
S = \sum_x \bigg[ \left(2d+m^2\right) \phi_x^*\phi_x 
 + \lambda\left( \phi_x^*\phi_x\right)^2
- \sum_{\nu=1}^4\left(  \phi_x^* e^{-\mu\delta_{\nu,4}} \phi_{x+\hat\nu} 
+ \phi_{x+\hat\nu}^* e^{\mu\delta_{\nu,4}} \phi_x \right)
\bigg].
\end{equation}
Splitting the field into its real and imaginary part, $\phi_x = \phi_{1,x} + i \phi_{2,x}$,
we can separate real and imaginary part of the action,
\begin{align}
S & = S^R+ i \sinh(\mu) S^I \nonumber \\
S^R
 &= \sum_x\bigg[ \frac{1}{2}\left(
 2d+m^2\right) \phi_{a,x}^2
 + \frac{\lambda}{4}\left(\phi_{a,x}^2\right)^2
 - \sum_{i=1}^3 \phi_{a, x}\phi_{a, x+\hat i}
 -\cosh(\mu) \ \phi_{a, x}\phi_{a, x+\hat 4}\bigg]
 \nonumber \\
S^I &= \sum_x \ \varepsilon_{ab}\phi_{a, x}\phi_{b, x+\hat 4}.
\end{align} \label{eq:bose_disc_exp}

It is worth noting that the phase quenched system differs from the system at zero chemical potential due to the presence of the $\cosh(\mu)$ term in the real part of the action. Throughout all our studies we have set the action parameters $\lambda=m=1.0$. This choice is motivated by the existence  of extensive literature results for this set of parameters. Moreover, it has been shown that, in this setup, numerical investigations results are well described by a mean-field evaluation.

\section{LLR Algorithm} \label{sect:llr}
The LLR (Linear Logarithmic Relaxation) algorithm provides a way to estimate the slope of the density of states by solving a non-linear stochastic equation. In the following we briefly review the method and our implementation.

We define a restricted and reweighed expectation value of a general operator $\mathcal{O}$ as following
\begin{equation}\label{eq:llr_exp_val}
\langle\langle {\cal O} \rangle\rangle_k(a) = \frac{1}{\mathcal{N}}
\int_{S^I_k - \Delta/2}^{S^I_k + \Delta/2} \rho(s) \ {\cal O}(s) \ e^{-a ( s - S^I_k)} \ \dd s
\end{equation}
for a given reweighing parameter $a$, and with $\rho(s)$ defined as in \eqref{eq:dos_def}. The normalisation factor $\mathcal{N}$ is defined as
\begin{equation}
\mathcal{N} =
\int_{S^I_k - \Delta/2}^{S^I_k + \Delta/2} \rho(s) \ e^{-a ( s - S^I_k)} \ \dd s.
\end{equation}

The heart of the LLR algorithm is the dynamical tuning of $a$, such that the reweighing factor $e^{-a ( s - S^I_k)}$ counterbalances the intrinsic density of states distribution of the system, resulting in an uniform sampling in a interval around $S^I_k$.

To achieve such a result we consider the specific observable ${\cal O}(s) = s-S^I_k$. As it has been shown in~\cite{Langfeld:2015fua}, in the limit of vanishing $\Delta$ the expectation value of this observable has a monotonous behaviour in $a$, and, more importantly, it vanishes when the corresponding value of $a$ coincides with the derivative of the DoS logarithm
\begin{equation}\label{eq:llr}
\langle\langle \Delta S^I \rangle\rangle_k(a)=
\langle\langle s-S^I_k \rangle\rangle(a) = 0 
\quad \Longleftrightarrow \quad a=a_k=
\eval{\dv{\ln (\rho(s))}{s}}_{s=S^I_k}
+ \order{\Delta^2}.
\end{equation}

Corrections of order $\Delta$ are not present due to the symmetry of the integrand function. 

To solve this implicit equation for $a$, we use two different techniques. Initially we use a Newton-Raphson method, generating a chain of reweighing factor $a_k^{(n)}$ according to
\begin{equation}
a^{(n+1)}_k =
a^{(n)}_k + \frac{\langle\langle \Delta S^I \rangle\rangle_k (a^{(n)}_k)}{\sigma^2(\Delta S^I, a^{(n)}_k)}.
\end{equation}
Approximating the variance of the distribution by
\begin{equation}
\sigma^2(\Delta S^I, a^{(n)}_k) \simeq \frac{\Delta^2}{12} + \order{\Delta^4},
\end{equation}
our actual update step can be written as
\begin{equation} \label{eq:nr}
a^{(n+1)}_k =
a^{(n)}_k + \frac{12 \ \langle\langle \Delta S^I \rangle\rangle_k (a^{(n)}_k)}{\ \Delta^2}.
\end{equation}
\begin{figure}[htb]
\begin{center}
\includegraphics[width=0.8\textwidth]{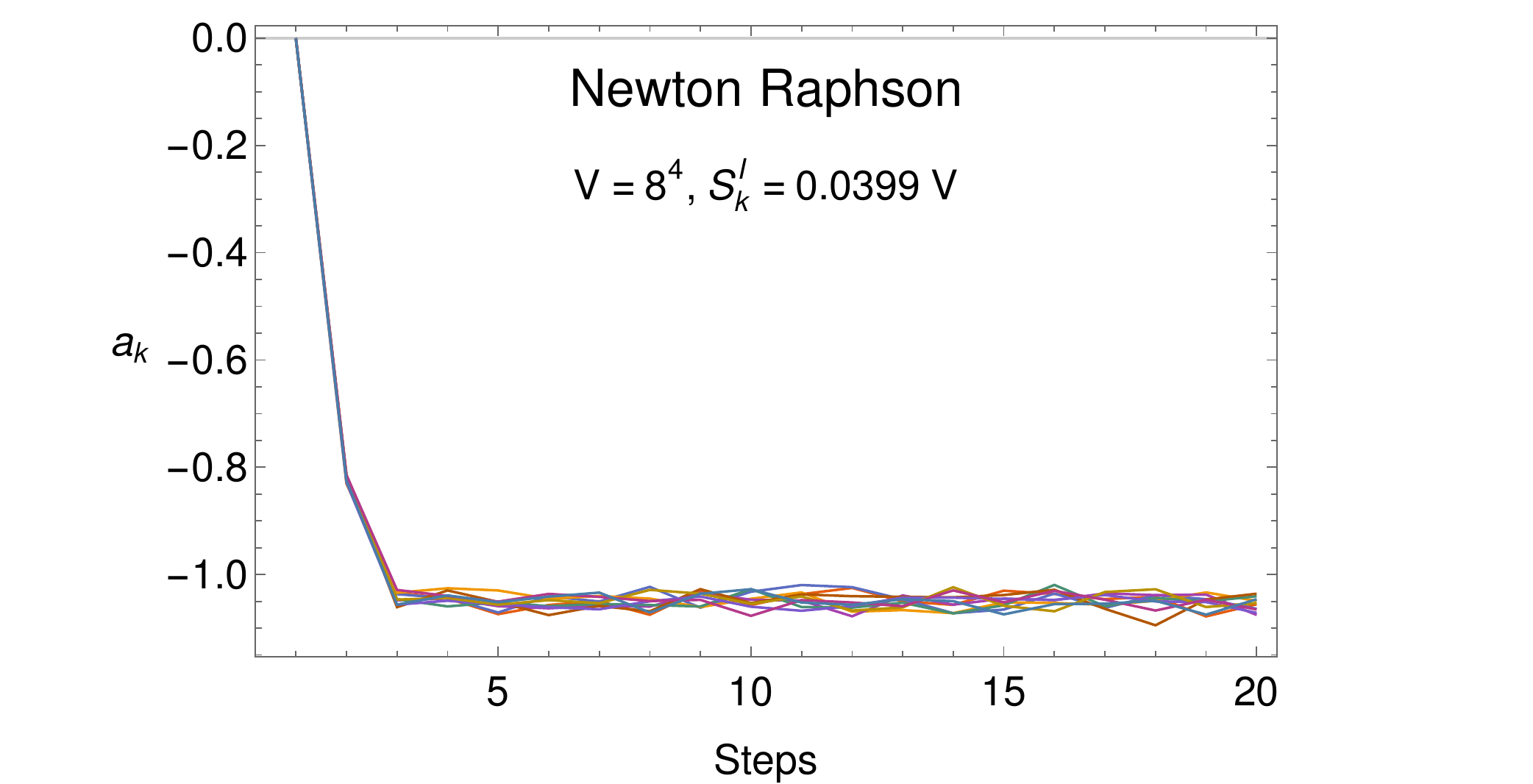}
\includegraphics[width=0.8\textwidth]{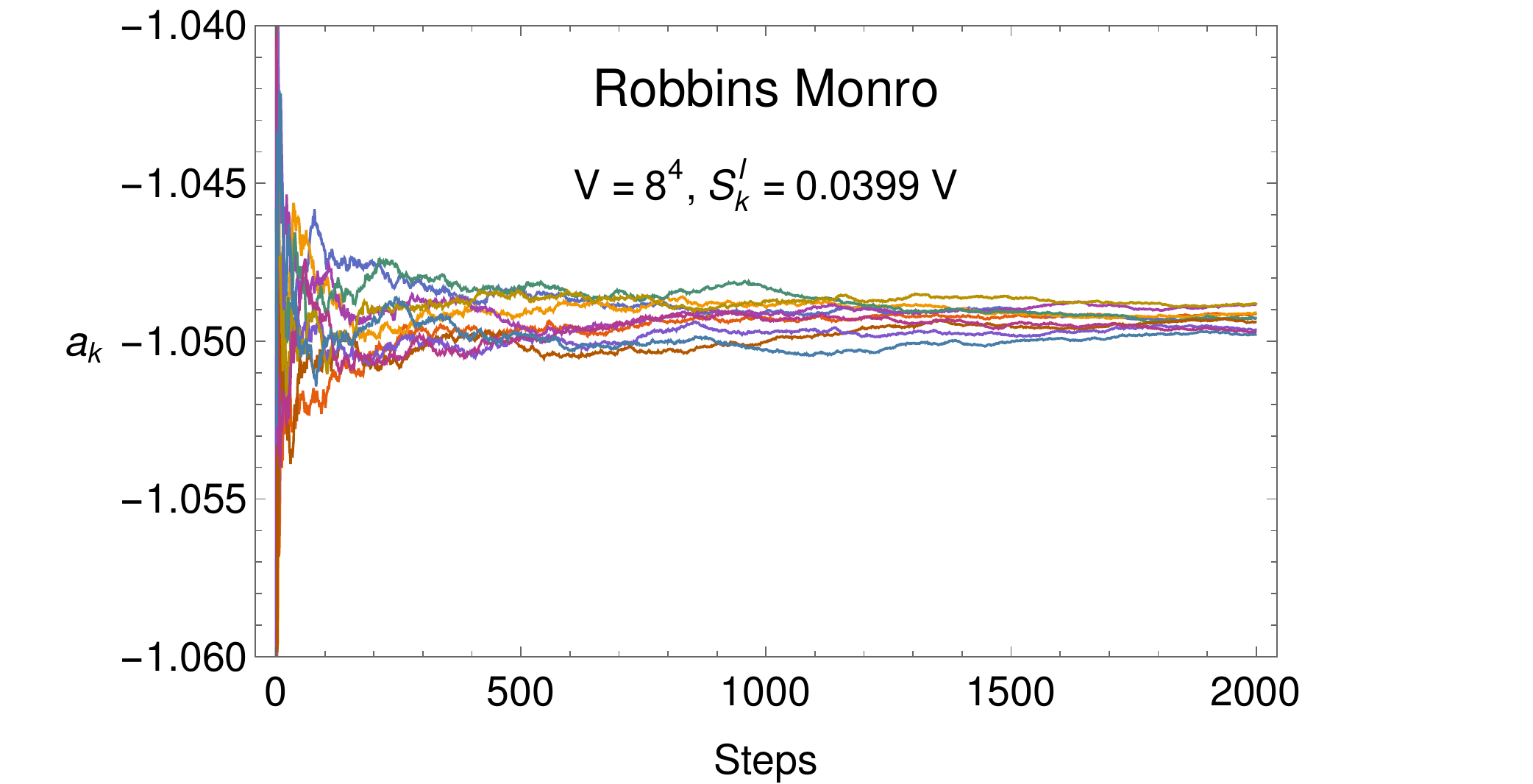} 
\end{center}
\caption{\textbf{Top}: evolution of the Newton-Raphson method for 10 independent simulations (replicas). A very rapid initial convergence towards the root of \eqref{eq:llr} and a subsequent non-converging oscillatory regime are clearly visible. \textbf{Bottom}: evolution of the Robbins-Monro stochastic root finding procedure for 10 independent simulations (replicas).}
\label{fig:NR_RM}
\end{figure}
As shown in Fig.~\ref{fig:NR_RM}, Newton-Raphson manages to approach the root extremely rapidly. However, due to the stochastic nature of Eq.~\eqref{eq:llr}, the statistical uncertainty intrinsic to the evaluation of $\langle\langle \Delta S^I \rangle\rangle_k(a_k)$ eventually prevents the Newton-Raphson method to converge to high level of precision. To overcome this issue, we employ the Robbins-Monro procedure~\cite{Robbins:1951}, applied to the determination of the $a_k$ once the once the Newton-Raphson method starts to oscillate around the solution. The Robbins-Monro method is based on iterative procedure
\begin{equation}
a^{(n+1)}_k =
a^{(n)}_k + c_n \frac{\langle\langle \Delta S^I \rangle\rangle_k (a^{(n)}_k)}{\sigma^2(\Delta S^I, a^{(n)}_k)}
\qquad
\sum_{n=0}^{\infty}c_n=\infty \ , \ \sum_{n=0}^{\infty}c_n^2 < \infty,
\end{equation}
where the conditions on the parameters $c_n$ ensure convergence to the correct root in the limit of $N_{RM}\rightarrow\infty$, where by $N_{RM}$ we indicate the total number of Robbins-Monro steps, even in presence of non white noise in the iteration estimator.
To maximise the speed of convergence, we choose $c_n=1/(n+1)$ to maximises the damping while respecting the bounds of the Robbins-Monro procedure leading to the sequence
\begin{equation} \label{eq:rm}
a^{(n+1)}_k =
a^{(n)}_k + \frac{1}{n+1}\frac{12 \ \langle\langle \Delta S^I \rangle\rangle_k (a^{(n)}_k)}{\ \Delta^2}.
\end{equation}

\begin{figure}[htb]
\centering
\includegraphics[width=0.8\textwidth]{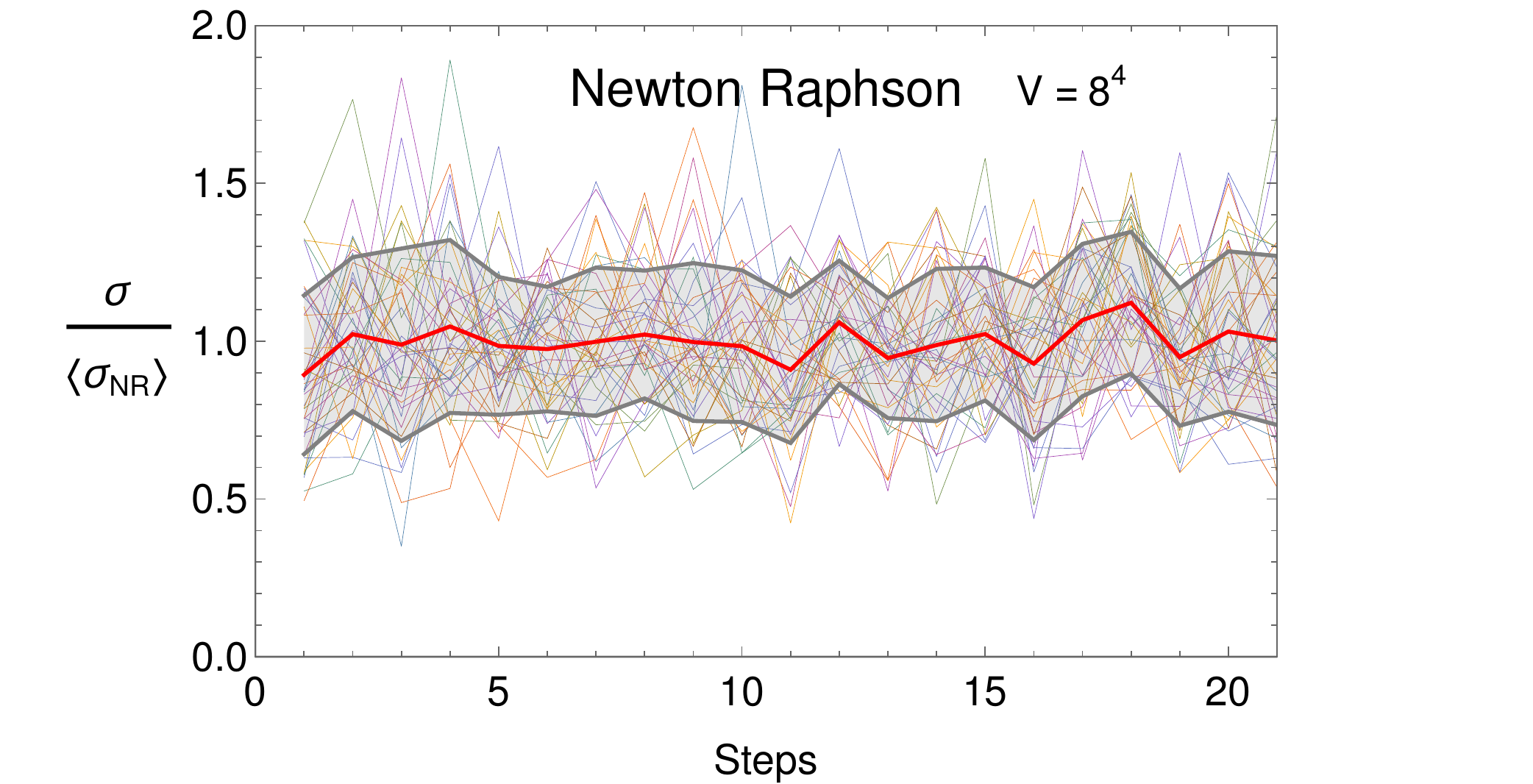} 
\includegraphics[width=0.8\textwidth]{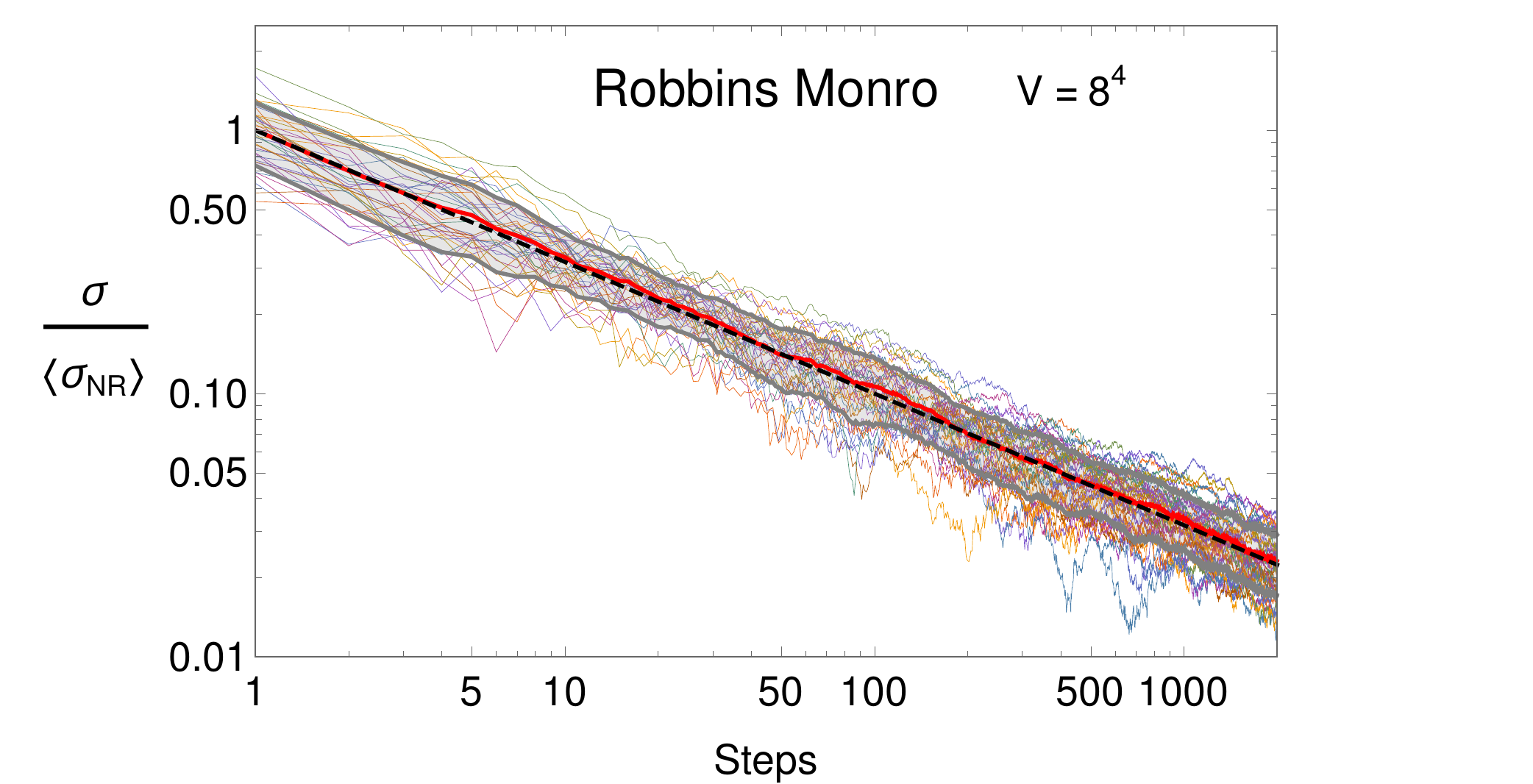} 
\caption{The plots show the standard deviation during the two different root finding procedures: \textbf{Top} Newton-Raphson, \textbf{Bottom} Robbins-Monro. Values of the standard deviation of independent simulations normalised to the mean value during the Newton-Raphson procedure are plotted against the number of root finding steps. Each thin coloured line represent a set of 10 independent simulations centred at the same value of $S_k$, while the red line is the mean of such values for different $S_k$. Plotted in grey, we show the $\pm 1 \sigma$ region and the dashed black line represent the theoretical best scaling of the standard deviation ($1/\sqrt{N_{RM}}$) for the Robbins-Monro procedure.}
\label{fig:NR_RM_var}
\end{figure}

Such a procedure converges in $L^2$ norm and hence in probability to the exact value, meaning that for each interval $\left[ S^I_k - \Delta/2,  \ S^I_k + \Delta/2 \right]$ the estimator $a^{(n)}_k$ is normally distributed around $a_k$ with variance scaling asymptotically as $1/N_{RM}$. We report in Figs.~\ref{fig:NR_RM}~and~\ref{fig:NR_RM_var} a detailed study of the convergence properties of the mean and standard deviation of the Robbins-Monro and Newton-Raphson algorithms, performed over various independent replicas reported as coloured lines in the plots. 

By applying this combined root finding procedure to all intervals, we can determine  $\dv*{\log\rho}{s}$ for an uniformly distributed set of $S^I_k$ values in the imaginary phase domain. In Sect.~\ref{sect:dos_reconstruction} we will discuss how to extend these results of the LLR procedure to the full domain of the DoS to fully reconstruct $\rho(s)$.

\section{LLR Intrinsic Bias} \label{sect:bias}
As it has been discussed in the previous section the LLR algorithm is exact for $\Delta \rightarrow 0$. However, this regime is unfavourable in numerical simulations as the Robbins-Monro step size scales as $\Delta^{-2}$ leading to huge jumps in the root finding procedure and a consequent really long convergence time. For this reason we are interested in studying the behaviour of the LLR algorithm when $\Delta$ is small, but not so much that the higher order correction to the DoS are negligible compared to the linear relaxation in the interval $\left[ S^I_k - \Delta/2, \ S^I_k+\Delta/2 \right]$. To do so we consider $\langle \langle \Delta S^I \rangle \rangle_k(a)$ with $a=a_k=\dv*{\log\rho}{s}\vert_{s=S^I_k}$, writing for ease of notation $\rho(s)=\exp(f(s))$ and including also higher order corrections
\begin{equation}
\langle\langle \Delta S^I \rangle\rangle_k(a=a_k) 
= \frac{1}{N}\int_{S^I_k-\Delta/2}^{S^I_k+\Delta/2} \ s \ e^{f(s)} \ e^{-a_k(s-S^I_k)} \ \dd s
= \frac{f^{(3)}(S^I_k)}{3!} \frac{\Delta^4}{80} + \mathcal{O}(\Delta^6).
\end{equation}
In the above the first order ($\order{\Delta^2}$) term vanishes for $a=a_k=f'(S^I_k)$, the second order ($\order{\Delta^3}$), linked to $f''(S^I_k)$, vanishes for symmetry of the integral, making the term of third order in the derivative the leading term. This term has the important characteristic of not depending on $a$, the reweighing parameter, meaning that it will introduce the same systematics in the Robbins-Monro procedure regardless of the distance from the root. In particular we can treat this term as an additive shift to Eq.~\eqref{eq:llr},
\begin{equation}
\langle\langle \Delta S^I \rangle\rangle_k(a \sim a_k) = \frac{\Delta^2}{12} \ (a - a_k) + \frac{f^{(3)}(S^I_k)}{3!} \frac{\Delta^4}{80} + \order{\Delta^6}.
\end{equation}
We can now evaluate what is the impact of this additive term on the reweighing parameter $a$ by solving the previous equation with the lhs set to zero. Obtaining,
\begin{equation}
\text{bias}=a_{\text{biased}}-a_k= \frac{f^{(3)}(S^I_k)}{40} \Delta^2 + \order{\Delta^4}.
\end{equation}
Therefore the bias will depend on two parameters: $f^{(3)}(S^I_k)$, the third derivative of the logarithm of the DoS that is system specific thus impossible to control a priori, and, as expected, $\Delta$, the interval width.
\begin{figure}[htb]
\centering
\includegraphics[width=0.7\textwidth]{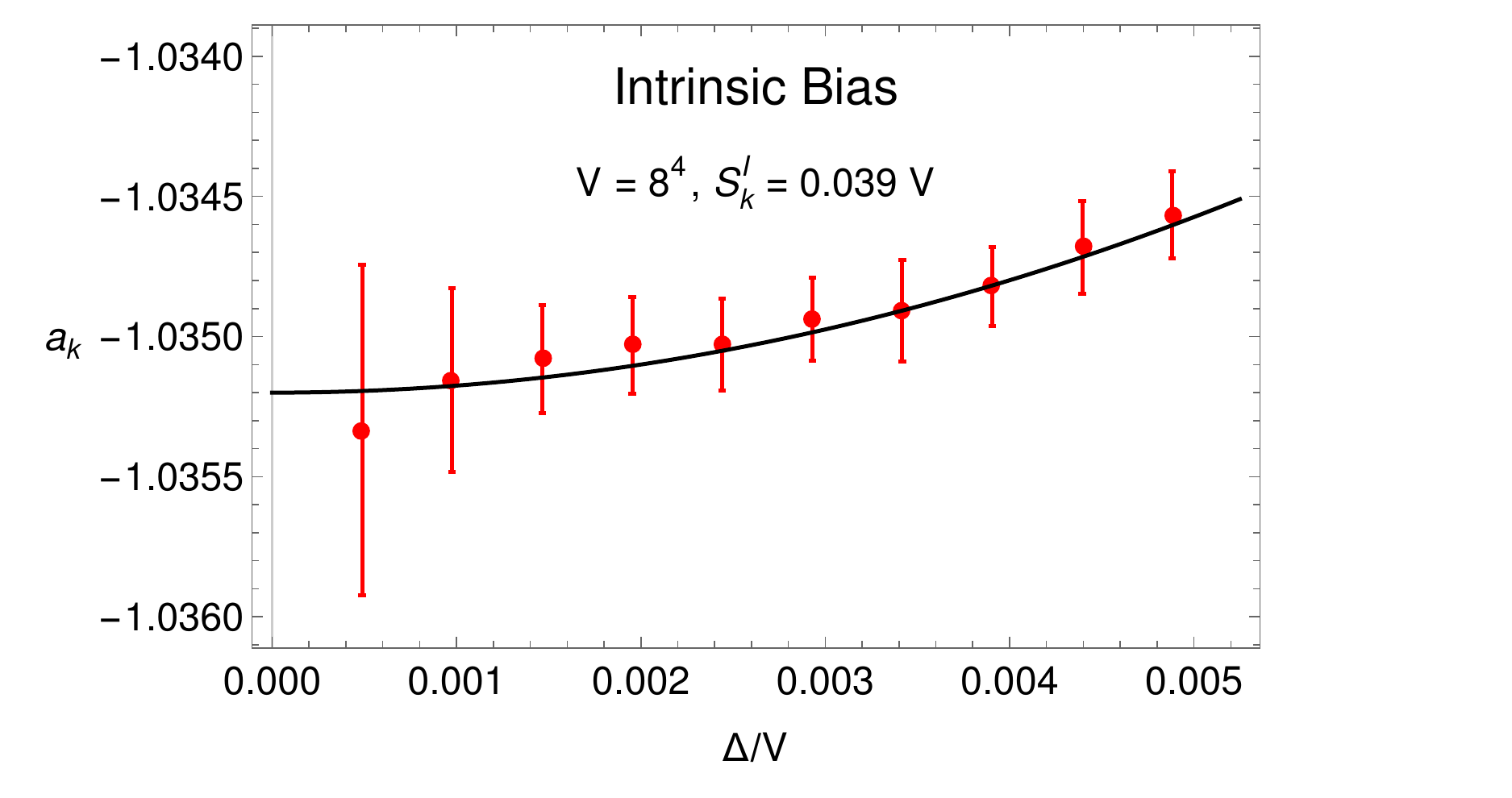} 
\caption{Simulation results (\textit{red points}) performed at different values of $\Delta$ compared to the predicted bias obtained by using the best fitting polynomial to estimate the third derivative of $\log \rho$. Both the bias effect and the increase in precision are clearly visible in the plot.}
\label{fig:bias}
\end{figure}
As shown in Fig. \ref{fig:bias} the effect of the bias is evident on the simulation results. Hence, it is possible to define a region for which the bias influence is negligible compared to the statistical uncertainty of independent simulations.

\section{DoS reconstruction techniques}\label{sect:dos_reconstruction}
We aim at a faithful reconstruction of the DoS of the system over the whole domain.
Thanks to the knowledge of $\dv*{\log\rho(s)}{s}$ many different reconstruction strategies can be formulated. In the following we will present two different choices of reconstruction and we will highlight the biases associated with each choice. 

Assuming the logarithmic derivative to be constant in each interval leads to the piecewise definition $\rho_{\text{pw}}(s)=\sum_k \hat{\rho}_k(s)$ with
\begin{equation} \label{eq:dos_pw}
\hat{\rho}_k(s) = C_k \ \exp\left( a_k ( s - S^I_k) \right),
\quad
s \ \in \ \left[ S^I_k - \Delta/2, \ S^I_k + \Delta/2 \right].
\end{equation}
and $\hat{\rho}_k(s)=0$ for $s$ outside the interval.
The parameters $C_k$ are chosen to ensure continuity 
\begin{equation} \label{eq:ck}
C_k= \exp \{a_k \ \Delta / 2 \} \ \prod_{i=0}^{k-1} \exp \{a_i \ \Delta\}.
\end{equation}

The LLR method achieves exponential error suppression, meaning that the relative error of $\rho(s)$ stays constant throughout the entire range of the imaginary action.
However, the piecewise approximation introduces a finite number of second order discontinuity in the imaginary action domain where the neighbouring exponentials are linked at the edge of the intervals. Such discontinuities will lead to precision issues in the evaluation of the oscillatory integral, Eq.~\eqref{eq:mv_phase}.

To overcome this limitation, we introduce our second reconstruction technique, the polynomial fitting ~\cite{Langfeld:2014nta} to substantially improve on the piecewise approximation.

In the polynomial fit approach the LLR results are fitted to a polynomial $p_l(s) = \sum_{i=0}^{l} \ c_i \ s^i$. An analytic integration of Eq. \eqref{eq:llr} allows to directly evaluate the $a_k$. Due to the symmetry properties of the DoS $\rho(s) = \rho(-s)$, only odd powers of $s$ enter into the polynomial, $p_l(s) = \sum_{i=1}^{l} \ c_{(2i-1)} \ s^{2i-1}$ and $c_i$ are determined  by fitting  our LLR results for $a_k$. 
The density of state resulting from the polynomial fitting can be expressed as
\begin{equation} \label{eq:dos_fit}
\rho_{\text{fit}(l)}(s) =
\exp{\int_0^s p_l(x) \ \dd x} =
\exp{\sum_{i=1}^l \frac{c_{(2i-1)}}{2i} \ s^{2i}},
\end{equation}
where we are normalising the DoS to have $\rho_{\text{fit}(l)}(0)=1$ as $p_l(0)=0$.
 
\begin{figure}[htb]
\centering
\includegraphics[width=0.8\textwidth]{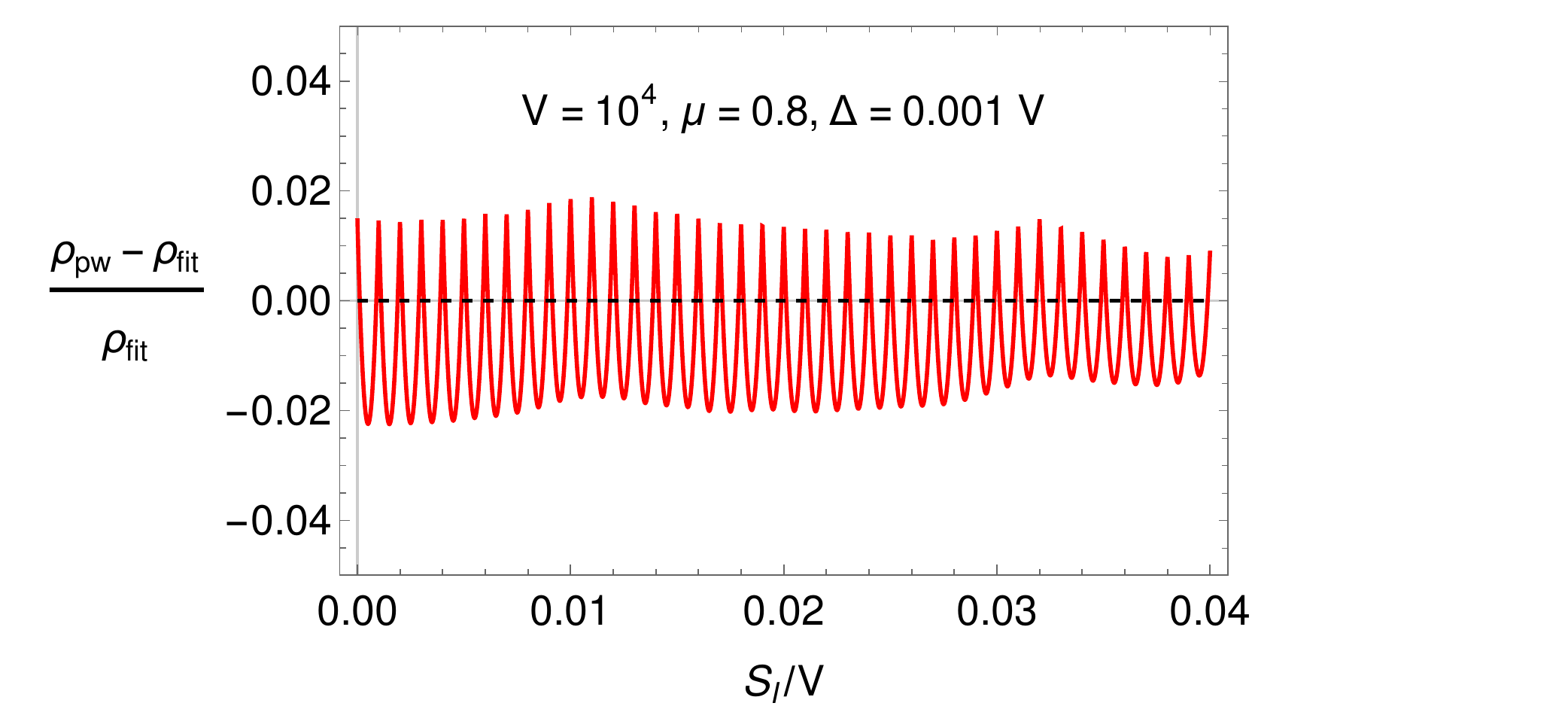} 
\caption{Comparison between the piecewise approximation of the DoS and the fitted approximation}
\label{fig:dos_pw_fit}
\end{figure}

As displayed in Fig.~\ref{fig:dos_pw_fit}, this approximation provides much smoother behaviour than the piecewise one, from which it shows bounded relative deviations. The jagged appearance of the plotted quantity results from the artefacts of the linear approximation involved in the reconstruction of the piecewise DoS.

To compare the values of the phase factor obtained with both approximations, we define the quantity
\begin{equation}\label{eq:avg_ph_part_int}
I (S^I_{\text{max}})  = \frac{\int_0^{S^I_{\text{max}}} ds \ \rho(s) \ \cos(\sinh (\mu) s)}{\int ds \ \rho(s)} \ , 
\end{equation}
This function can be evaluated with both the piecewise or fitted definition of the DoS. In addition, $I(S^I_{\text{max}})$ is related to the expectation value of the phase factor by
\begin{equation}
\frac{1}{M} \sum_{m = 1}^M \left( \lim_{S^I_{\text{max}} \to \infty} I_{m} (S^I_{\text{max}})    \right) = \langle e^{i \varphi} \rangle_{pq} \ , 
\end{equation}
where $M$ is the size of an ensemble of gaussianly distributed realisations of the $a_k$.

\begin{figure}[htb]
\centering
\includegraphics[width=0.7\textwidth]{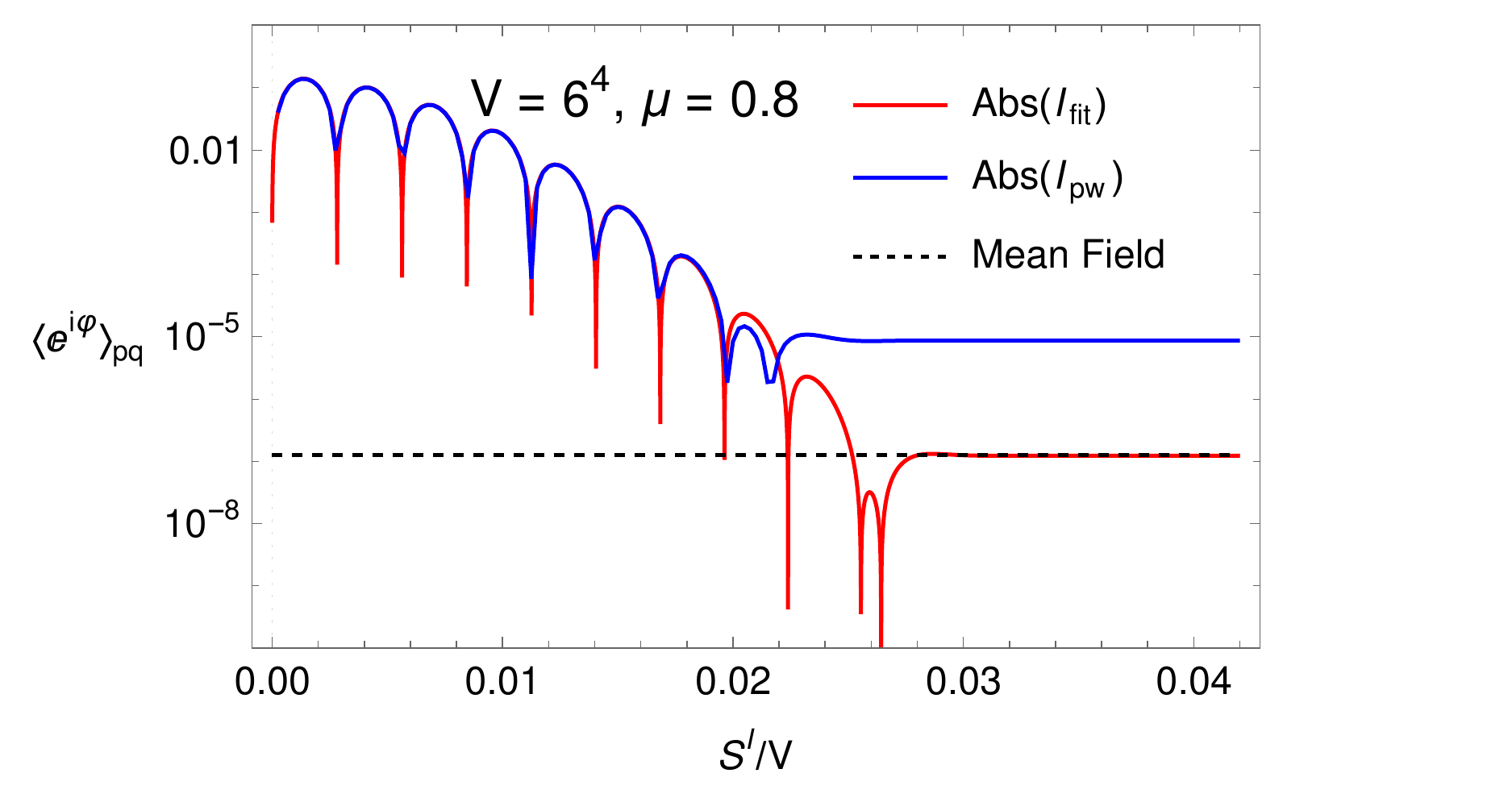}
\includegraphics[width=0.7\textwidth]{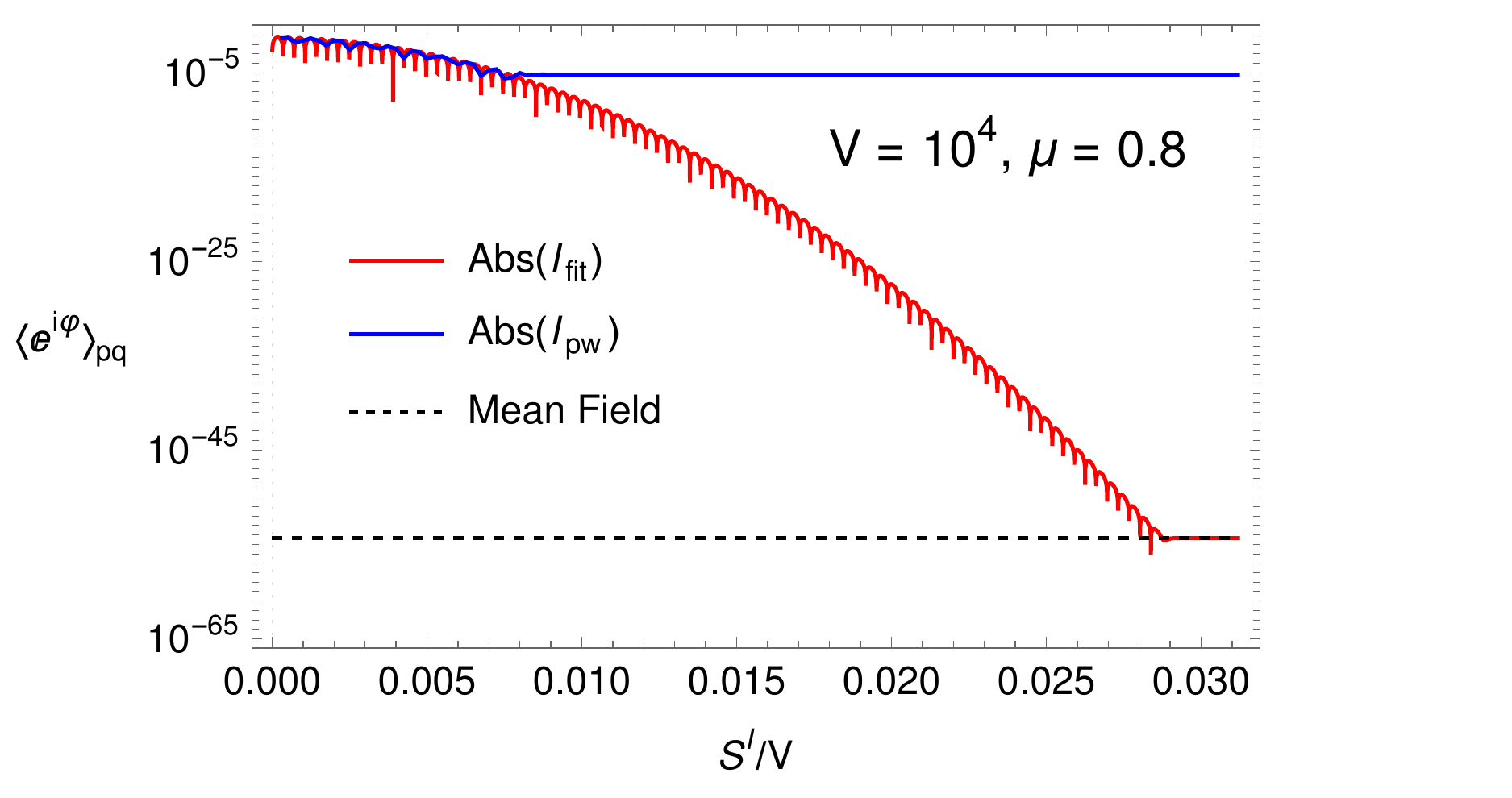}
\caption{Partially integrated phase factor, Eq.~\eqref{eq:avg_ph_part_int}, as a function of the upper integration limit for volumes $V=6^4$ and $10^4$ at $\mu=0.8$. Here, we plot the absolute value of the partially integrated phase factor on a logarithmic scale for ease of visualisation we are plotting the $abs()$ of the phase factor, such a choice affects only the region for which the integral has not yet converged.}
\label{fig:pw_fit}
\end{figure}

In Fig.~\ref{fig:pw_fit} we compare the absolute values of the partially integrated phase factor as a function of $S^I_{\text{max}}$ for both approximations for two different volumes $V = 6^4, \ 10^4$ at $\mu = 0.8$ and a particular realisation of the $a_k$. Our results show that using the piecewise approximation generates much larger fluctuations than the polynomial interpolation: for the $10^4$ the relative amplitude of the fluctuations is of around 45 orders of magnitude. While, when averaged over multiple realizations the $a_k$ coefficients, both definitions give compatible results, only the polynomial interpolation provides a value that is accurately different from zero within the precision of the calculation and hence allows us to detect the severe cancellations generated by the sign problem.  Indeed the piecewise approximation fails to achieve sufficient precision as the integration generates an intrinsic error of $\order{\Delta^2}$ for each interval (due to the correction to the linear approximation neglected in this procedure). When the sign problem gets exponentially hard, an exponentially large number of small intervals should be taken into account to achieve the required precision in order to suppress the intrinsic error. On the other hand, the polynomial approximated DoS seems to show no difficulty at obtaining an accurate result broadly compatible with the mean-field calculation also for the harder $V = 10^4$ case.

\begin{figure}[htb]
\centering
\includegraphics[width=0.8\textwidth]{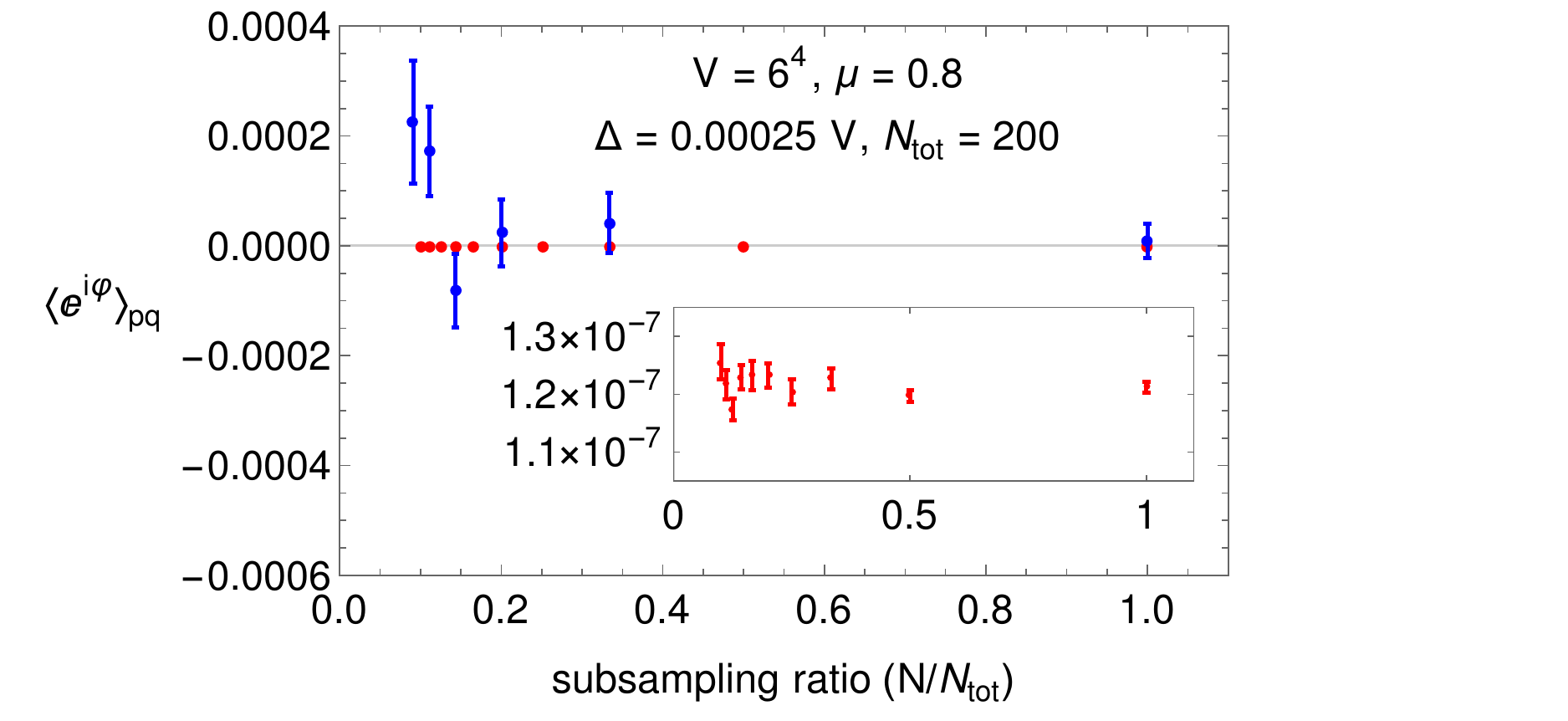}
\caption{Results of the integration of the phase factor for different values of the subsampling ratio for a $V=6^4$ lattice at $\mu=0.8$. In the figure are plotted the results of the fitted approach integration (red) and those of the piecewise one (blue). The inset shows the remarkable level of precision obtainable with the fitted approach.}
\label{fig:subsmpl}
\end{figure}

\begin{figure}[htb]
\centering
\includegraphics[width=0.8\textwidth]{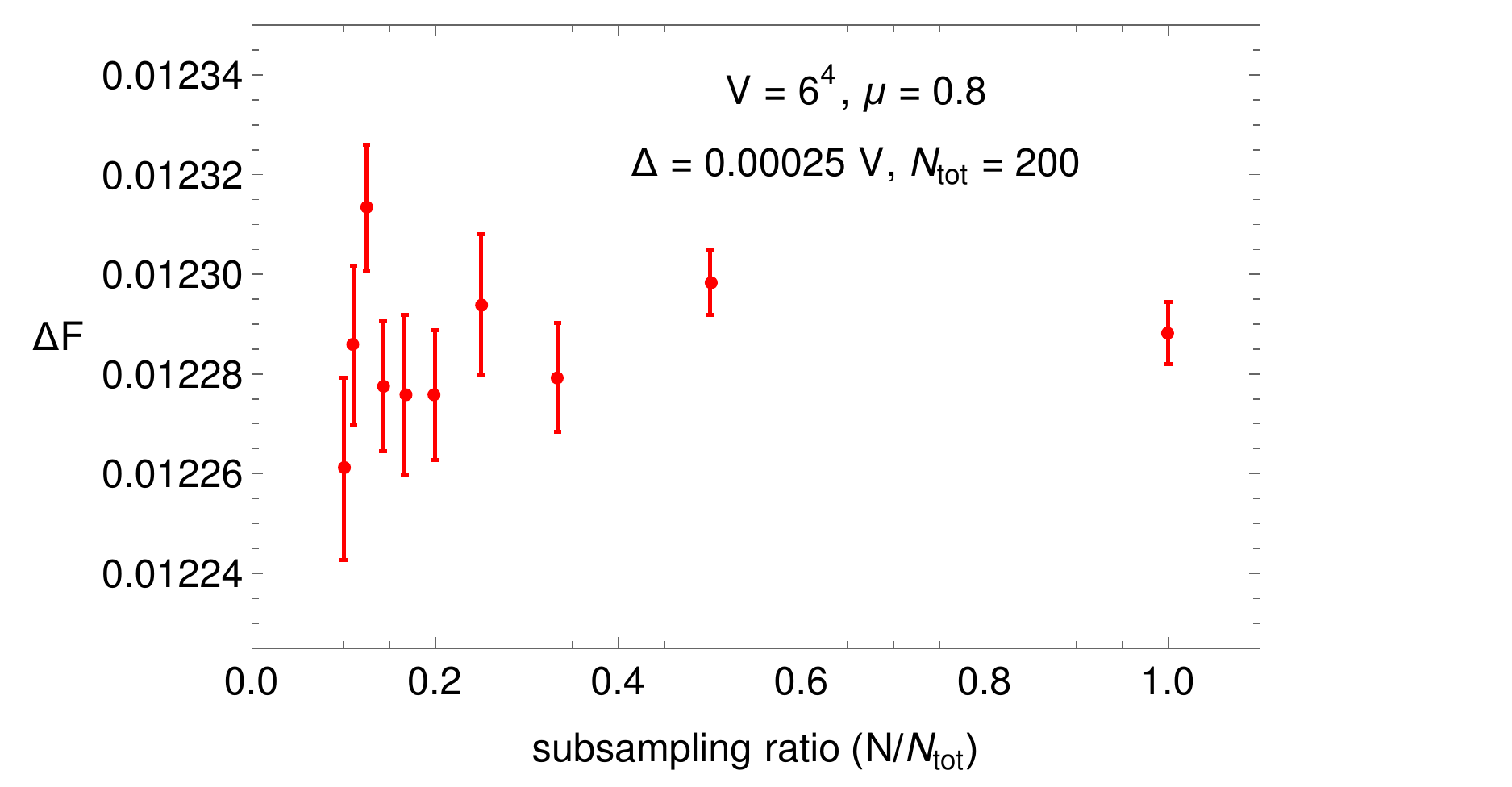}
\caption{Values of the free energy difference obtained with the data shown in Fig. \ref{fig:subsmpl} only for the fitted approach.}
\label{fig:subsmpl_df_fit}
\end{figure}

The different precision obtained by the two methods can be analysed further by computing the phase average. In order to check the convergence of the result, we study the latter quantity for different coarse-graining of the $a_k$, obtained by taking subsets with different spacing between consecutive values ({\em subsampling}). In Fig.~\ref{fig:subsmpl},  we contrast the level of precision on $\langle e^{i \varphi} \rangle_{pq}$ obtained with the two methods as the spacing between two central values of the $S_I$ used to calculate the $a_k$ varies. For our finest determinations (i.e., for $N/N_{tot}$ approaching the value of one, which means that all the values we have determined are used in the reconstruction), the data converge to an asymptotic value, from which they deviate for coarser spacing ($N/N_{tot} \ll 1$). However, while the polynomial fit provides a reliably accurate determination of the average sign, the use of the piecewise interpolation generates a statistical error that makes the result compatible with zero. Fig.~\ref{fig:subsmpl_df_fit} shows the quality of the determination of the free energy difference $\Delta F$ corresponding to the polynomial fit. 

In addition to the polynomial fit, we have performed other interpolations to understand the regime of validity of the results. In particular, we have interpolated the $a_k$ using
\begin{itemize}
\item {Expansions in  an $L^2$ basis} (in particular, using Hermite functions);
\item {Continuous local fits (loess/lowess)}, whereby a local low-order polynomial fit is convoluted with localised weight functions;
\item {Gaussian processes}, which use a multi-variate Gaussian a priori ansatz for modelling the distribution of correlations among $n$-tuples of observations. 
\item {\em Pad\'e approximants} of various order.
\end{itemize}  
Somehow surprisingly, all these methods produced results that were less accurate than the simple (and {\em a priori} simplistic) polynomial interpolator, basically failing at disentangling a non-zero average phase from the noise when a hard sign problem is present. The different reasons for the observed failures are instructive:
\begin{itemize}
\item The $L^2$ expansion converges slowly, hence requiring a high number of terms or equivalently a high number of fitted parameters, which results in detectable overfitting (we will discuss overfitting in the context of the polynomial interpolation below); 
\item Continuous local fits are too sensitive to the locally projecting functions, generating noise at a frequency that is roughly the inverse of the amplitude of the window on which one performs the projection;
\item Gaussian processes presented localised high-frequency oscillations that also resulted in noisy measurements for the Fourier transform.
\end{itemize}
Amongst those methods, perhaps the most surprising failure is associated to Pad\'e approximants, which in general are expected to converge faster than polynomial interpolations. The better outcomes obtained with the latter may indicate that the variation of the $a_k$ with $S_I$ is indeed described by a (near-)polynomial function.  We believe that this information is physically relevant for understanding the system. A possible explanation of the success of the polynomial interpolation may be inferred from the behaviour of the $a_k$ as a function of the relevant observable (e.g., the energy) for systems at zero density, whereby higher power contributions are suppressed by powers of the volume~\cite{Langfeld:2015fua}. It is possible that this {\em local} property holds on a wider scale.

Having shown that, unlike other choices, the polynomial fitting approach of the DoS allows us to achieves a high level of precision for the determination of the average sign, we shall now address the numerical stability of the approach with respect to the order of the chosen polynomial and estimate possible systematics related to the determination of the maximum power of $s$ appearing in the polynomial. 

\section{Fit Validation} \label{sect:fit_validation}
Concerning the stability of the polynomial fit, the choice of the polynomial order $l$ is of course of crucial importance. In this section we will illustrate how to ensure that the functional form choice avoids the two most likely source of systematics, \textit{under}-fitting and \textit{over}-fitting.

\subsection*{Under-fitting}
Under-fitting happens when the proposed polynomial is too simple (the order is too low) to represent all the features of the data. In this case a $\chi^2$ analysis of the fit residuals is able to pin down the minimum number of polynomial coefficients needed to describe the LLR results.
\begin{figure}[htb]
\centering
\includegraphics[width=0.6\textwidth]{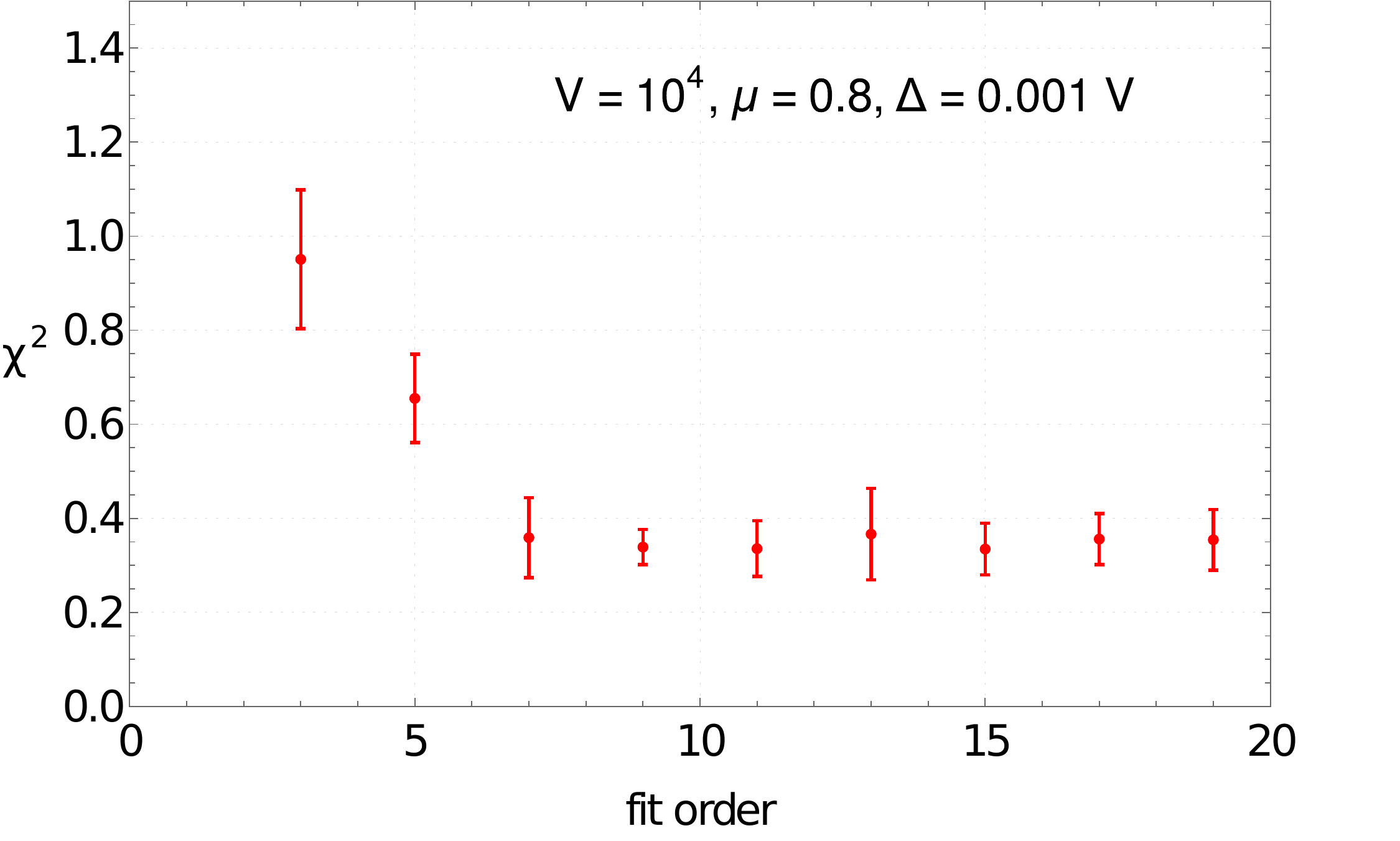}
\caption{$\chi^2$ values resulting from a bootstrap analysis vs. the order of the fit, a clear plateau is visible starting from $l=7$.}
\label{fig:chi2}
\end{figure}
As shown in Fig.~\ref{fig:chi2}, the $\chi^2$ value decreases while increasing the order of the fitted polynomial. For the data considered at $l=7$ we start to see a plateau forming. After the onset of the plateau, due to the statistical uncertainty of our data, higher order polynomials will start to pick up the statistical noise rather than improve the approximation. For this reason one could be tempted to choose simply the smallest order in accordance with the Occam's razor principle. Instead, we evaluate \eqref{eq:mv_phase} for a various choices of the polynomial fit order. If we see a plateau also in the expectation values of the phase factor we will accept the results, otherwise we will reject the result and proceed to increase the precision of the simulation.

\subsection*{Over-fitting}
The other way in which the fitting process could introduce a systematic error is {\em over-fitting}, when the polynomial order is so high that the fitted function will start to introduce noise not related to the statistical uncertainty of the fitted data. To control the over-fitting we are going to study the expectation values of the second order derivative of $\log\rho$ with the intent of comparing the numerical values with the derivative of the fitted polynomial. 

\begin{figure}[htb]
\centering
\includegraphics[width=0.7\textwidth]{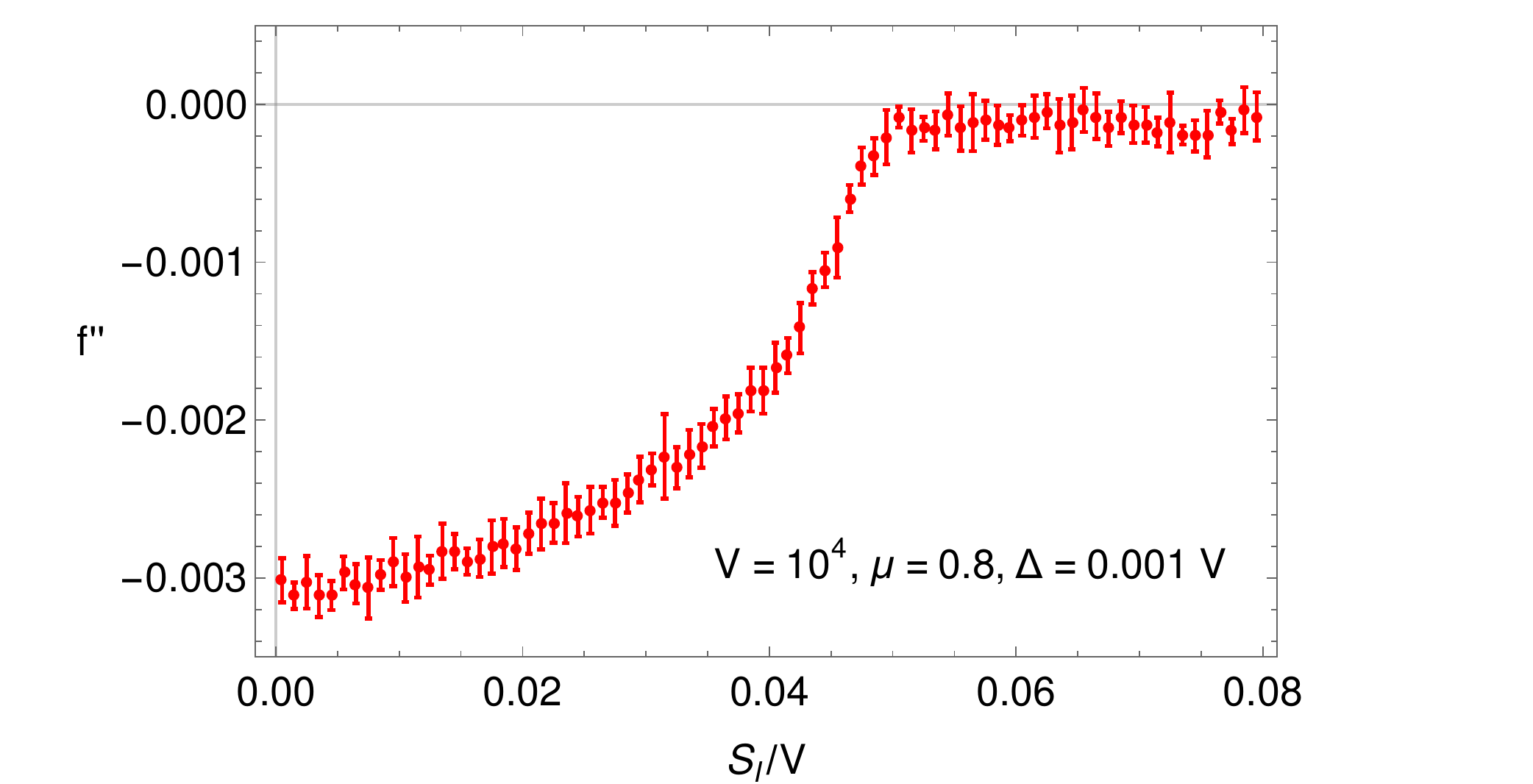}
\caption{Second logarithmic derivative $f''(S^I_k)$ obtained from the evaluation of $\langle\langle (\Delta S^I)^2 \rangle\rangle$ from a simulation at lattice volume of $10^4$ at $\mu=0.8$.}
\label{fig:f2}
\end{figure}

\begin{figure}[htb]
\centering
\includegraphics[width=0.7\textwidth]{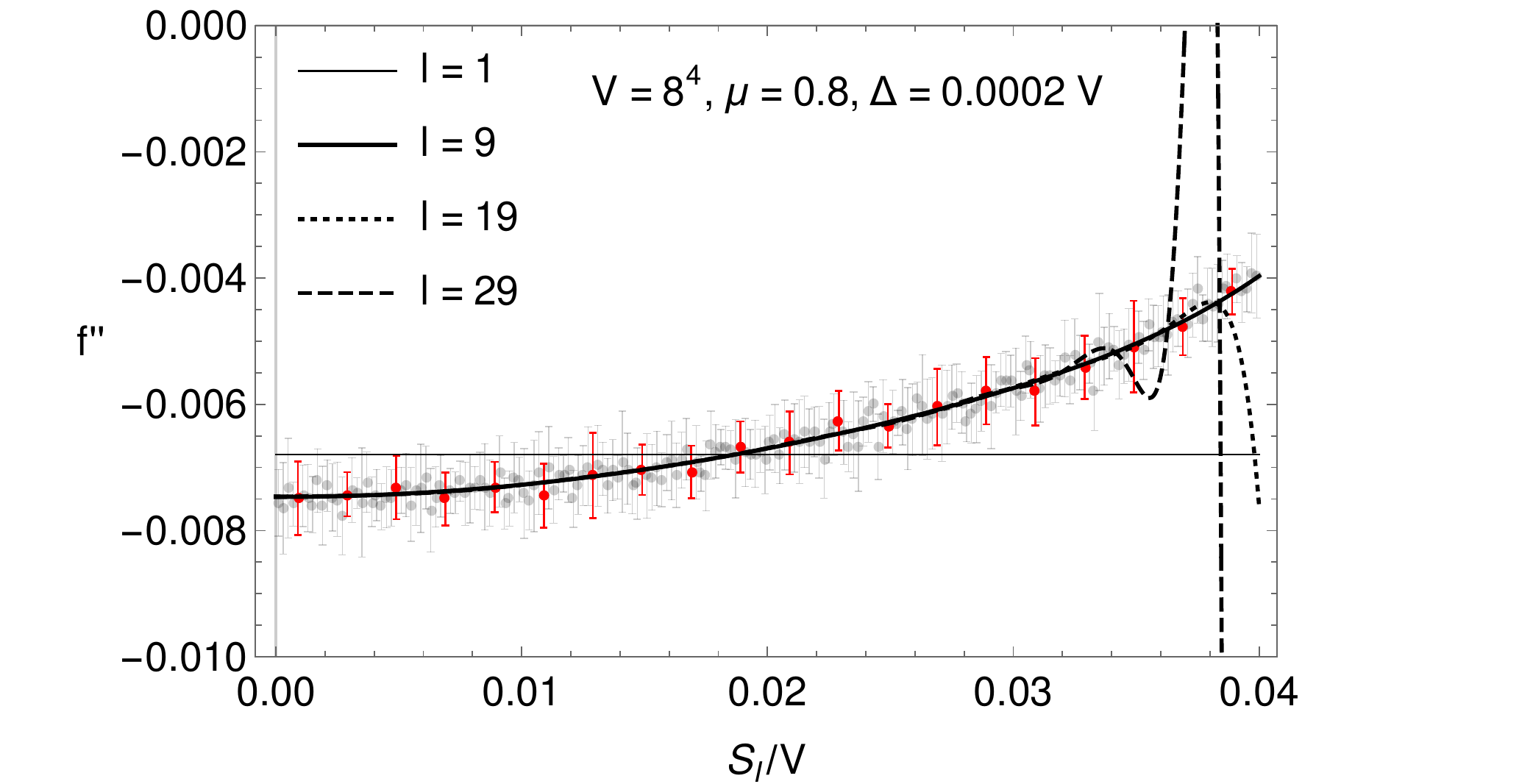}
\includegraphics[width=0.7\textwidth]{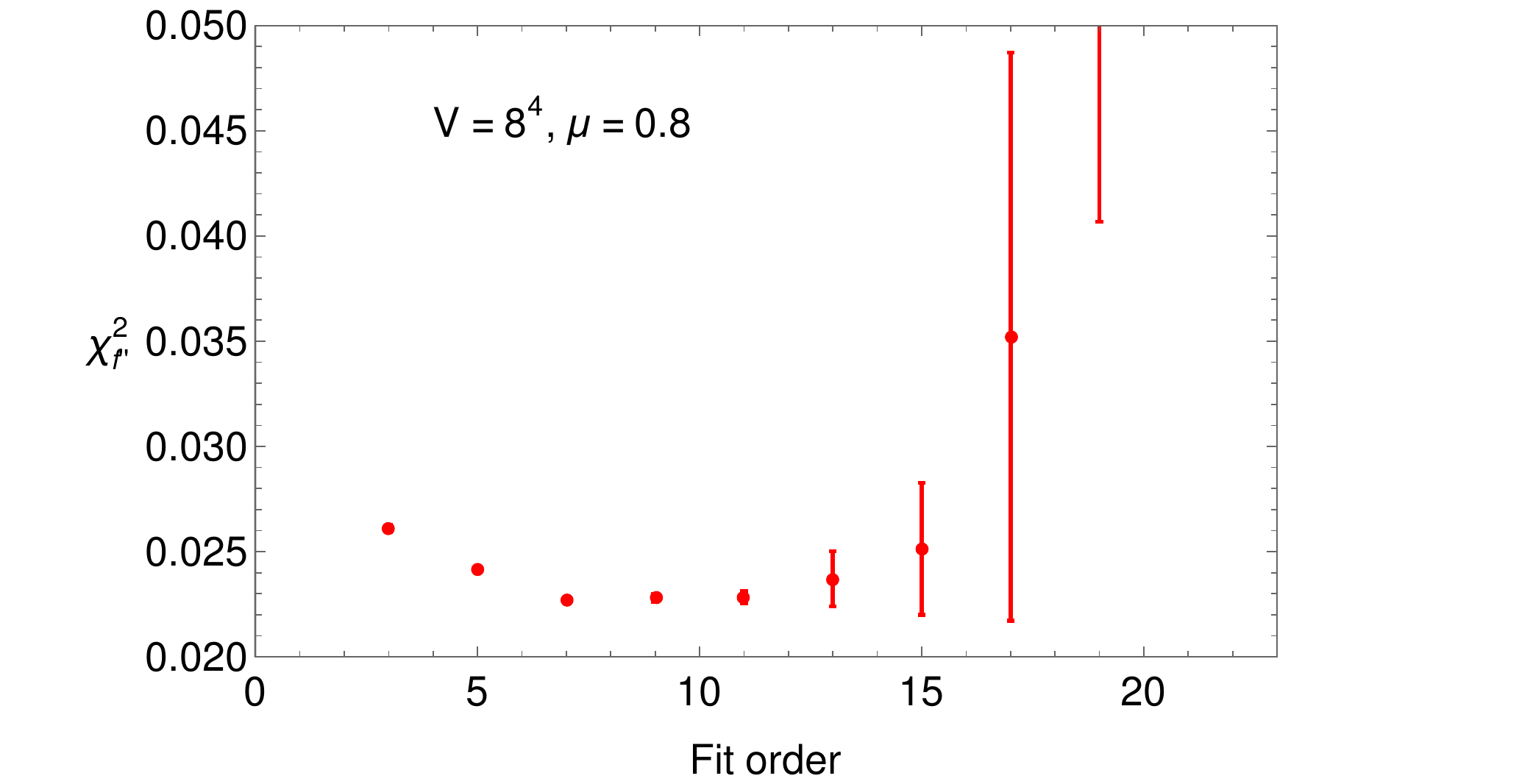}
\caption{\textbf{Top}: $f''(s)$ values obtained as result of our simulation (dots) and as the derivative of the fitted polynomials (lines). The highlighted red dots are those used in the $\chi^2$ analysis. \textbf{Bottom}: $\chi^2$ analysis for the second derivative.}
\label{fig:df2}
\end{figure}

The second derivative of $\log\rho$ can be determined numerically by evaluating restricted and reweighed expectation values \eqref{eq:llr_exp_val}.
We start by determining $\langle\langle (\Delta S^I)^2 \rangle\rangle_k$ with $a=a_k$, where we are writing $\rho(s) = \exp{f(s)}$,
\begin{equation} \label{eq:ds2}
\langle\langle (\Delta S^I)^2 \rangle\rangle_k (a=a_k) =
\frac{\Delta ^2}{12}+\frac{f''(S^I_k)}{360} \ \Delta^4 + \order{\Delta^6}
\end{equation}
from which it is possible to obtain the value of the second derivative as
\begin{equation} \label{eq:d2f}
f''(S^I_k) = \frac{360}{\Delta^4}\left( \langle\langle (\Delta S^I)^2 \rangle\rangle_k - \frac{\Delta^2}{12} \right) + \mathcal{O}(\Delta^2).
\end{equation}
This quantity is measurable to an acceptable level of statistical relevance in our simulations, as shown in Fig.~\ref{fig:f2}, despite coming from an evaluation of the second moment of the distribution.
Rather than using the second derivative directly in the fitting procedure we look at how well the polynomial fit of the $a_k$ describes this quantity (i.e. we perform an {\em a posterior} validation of the functional form of fit). We compare $f''(s)$ with the derivative of the polynomial fit $p_l'$ by defining a $\chi^2$-like function
\begin{equation}
\chi^2_{f''} =\frac{1}{N} \sqrt{\sum_{i=1}^{N} \left( p'_l(s_i) - f''(s_i) \right)^2/\sigma_{f''(s_i)}^2}.
\end{equation}
To illustrate this principle we consider a sub-sampling of our data, so that the effects of the over-fitting are evident even at lower order of the fitting polynomial. The results of this analysis are shown in Fig.~\ref{fig:df2}, where three different behaviours are visible: for small orders of the polynomial fit($l=1\sim5$) the discrepancies are big as the polynomial is not a good approximation of the original data; in the middle section ($l=7\sim11$) we see a plateau indicating that in this region the fit not only approximates well $f'(s)$, but also its derivative; for high orders ($l\geqslant13$) we see  a clear indication of over-fitting, as, while the $\chi^2$ of the fit of the $a_k$ would still be on the plateau, $\chi^2_{f''}$ shows that the fitted function does not approximate well the numerical derivative.
This gives us a quantitative indication of whether the chosen functional form is over-fitting the data. When using the full set of data no sign of over-fitting has been found in any of our analysis.

\section{Bias optimised simulations}\label{sect:bias_opt}
The following scheme ensures a bias free and performance optimised simulation:
\begin{itemize}
\item Run a low precision simulation (fewer Monte Carlo samplings ($N_{MC}$) as well as Robbins-Monro steps ($N_{RM}$)) with a small and constant $\Delta$ for  each interval, extract the values of the $a_k$, and use those to estimate the bias over the complex action range taken into consideration.
\item Scale the simulation parameters ($N_{MC}$, $N_{RM}$ and $\Delta$) so that $\text{bias} \ll \sigma_{a_k}$.  We use the known scalings $\text{bias} \propto \Delta^{2}$ and $\sigma_{a_k} \propto $ $(\Delta \cdot \sqrt{N_{MC} \cdot N_{RM}})^{-1}$, and the fact that the simulation runtime is proportional to  $N_{MC} \cdot N_{RM}$. 
\item With the scaled parameters run a high precision simulation, the results of which will be used to rebuild the DoS.
\item Finally, using the high precision results double check that the bias is negligible in comparison to the statistical noise of the results.
\end{itemize}

\section{Results}\label{sect:res}

Following the scheme described in the previous sections we have been able to obtain the $a_k$ estimates for a wide range of values in the chemical potential, ranging from $\mu=0$ to $2.0$, and volumes ranging from $4^4$ to $16^4$. The typical values of the simulation parameters are reported in Tab~\ref{tab:1}. 

\begin{figure}[htb]
\centering
\includegraphics[width=0.7\textwidth]{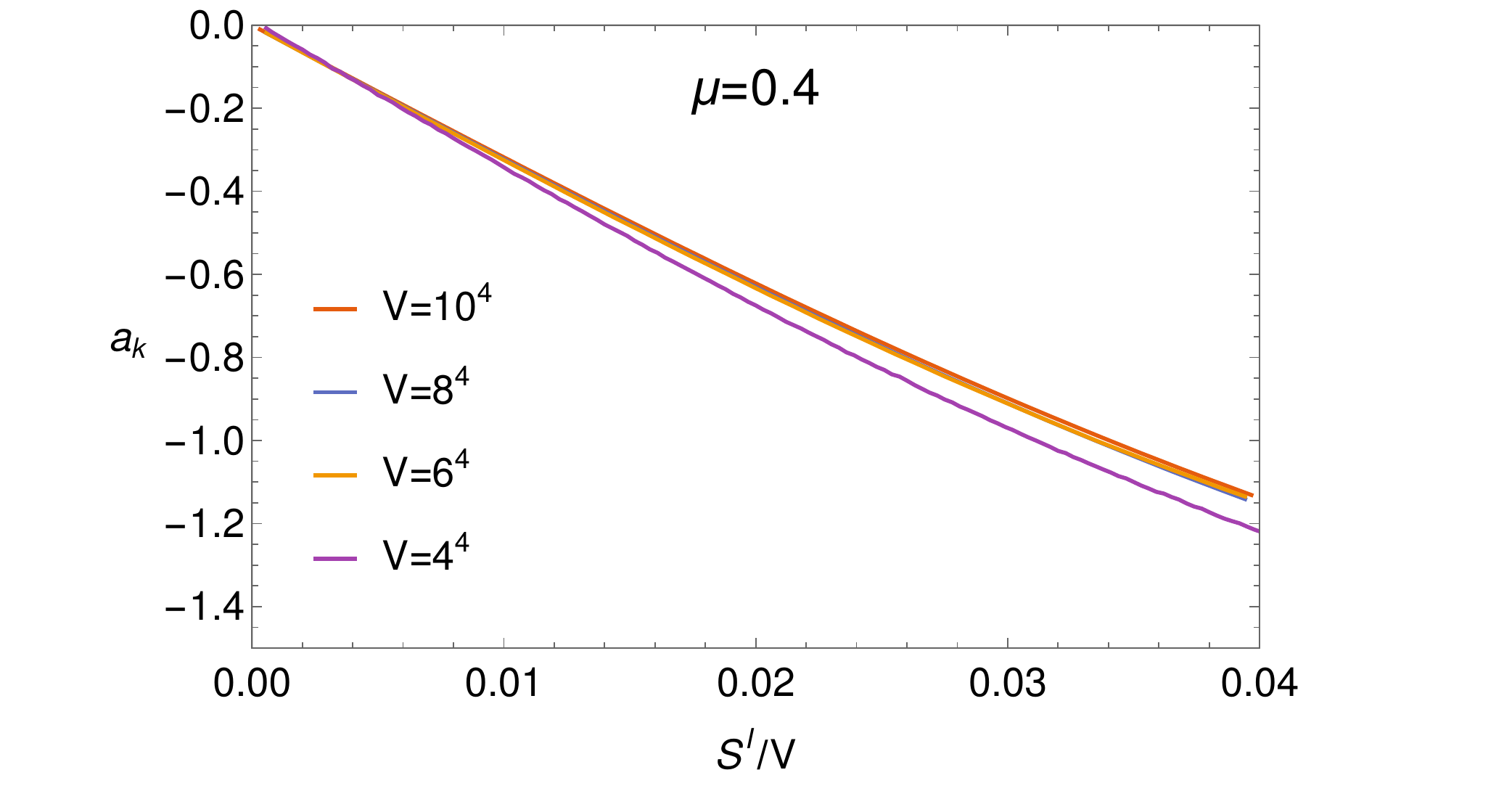}
\includegraphics[width=0.7\textwidth]{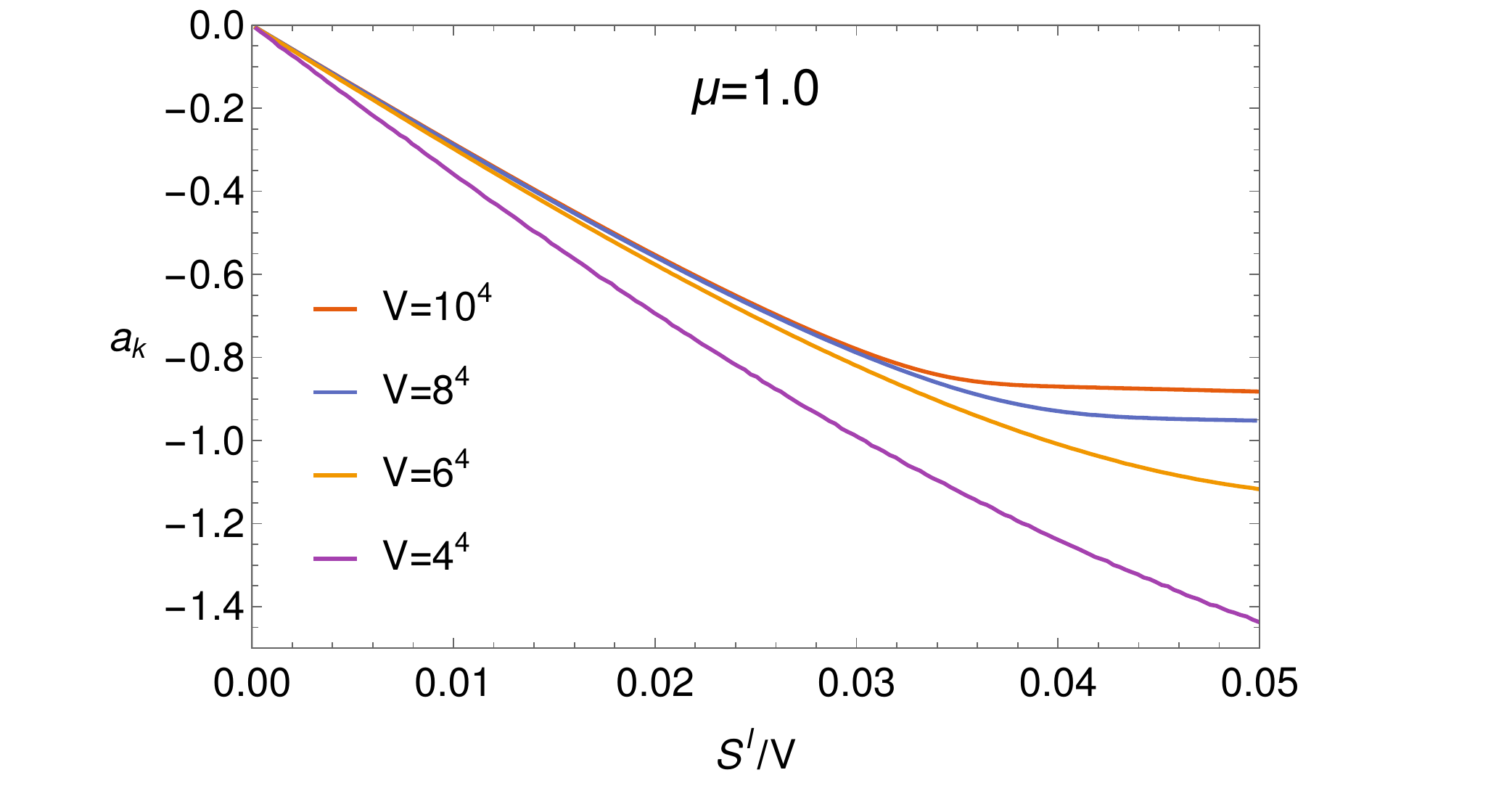}
\includegraphics[width=0.7\textwidth]{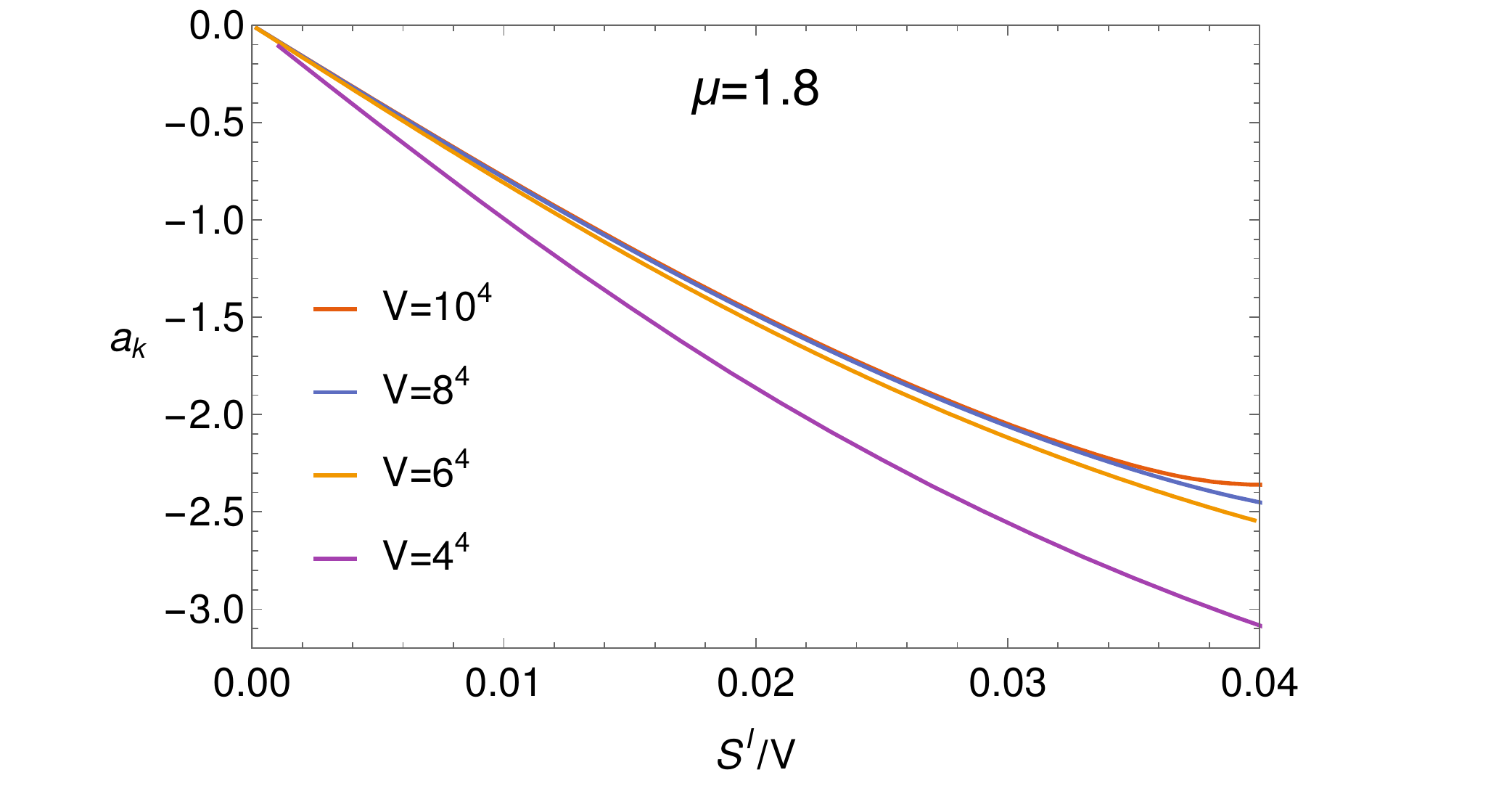}
\caption{Estimates of the $a_k$ for the relativistic Bose gas for different values of the imaginary part of the action at chemical potential $\mu=0.4$ (top), $\mu=1.0$ (middle) and $\mu=1.8$ (bottom) for different volumes.}
\label{fig:ak_dos}
\end{figure}

A representative set of results of this evaluation has been reported in Fig.~\ref{fig:ak_dos}. A general feature of the $a_k$ as a function of $S^I$ is the appearance of a sharp change of behaviour for large $S^I$ when $\mu$ is close to a critical value, the location of the inflection point decreasing for larger volumes. If the inflection point is present in the interval of imaginary action relevant for our numerical integration the fitting procedure defined in the previous section fails to converge. As a consequence, we can estimate the free energy only if the above change of behaviour does not occur, hence for large volumes we can do the integral only outside a region around the critical $\mu$. 

\begin{table}[htb]
\centering
\begin{tabular}{ccccccc}
\multicolumn{1}{c|}{V}      & $N_{NR}$ & $N_{RM}$ & $N_{MC}$ & $N_{S^I_k}$ & $N_{rep}$ \\ \cline{1-6}
\multicolumn{1}{c|}{$4^4$}  & 10       & 1000     & 2000     & 40          & 10        \\
\multicolumn{1}{c|}{$6^4$}  & 10       & 1000     & 2000     & 80          & 10        \\
\multicolumn{1}{c|}{$8^4$}  & 20       & 2000     & 2000     & 160         & 10        \\
\multicolumn{1}{c|}{$10^4$} & 50       & 2000     & 2000     & 200         & 10        \\
\multicolumn{1}{c|}{$16^4$} & 50       & 2000     & 2000     & 300         & 10        \\
\end{tabular}
\caption{Typical simulation parameters. $N_{NR}$: Newton-Raphson steps; $N_{RM}$: Robbins-Monro steps; $N_{MC}$ Monte Carlo samples for each step; $N_{S^I_k}$: intervals taken into consideration; $N_{rep}$: independent replicas.\label{tab:1}}
\end{table}

\subsection*{Phase structure away from criticality}
\begin{figure}[htb]
\centering
\includegraphics[width=0.7\textwidth]{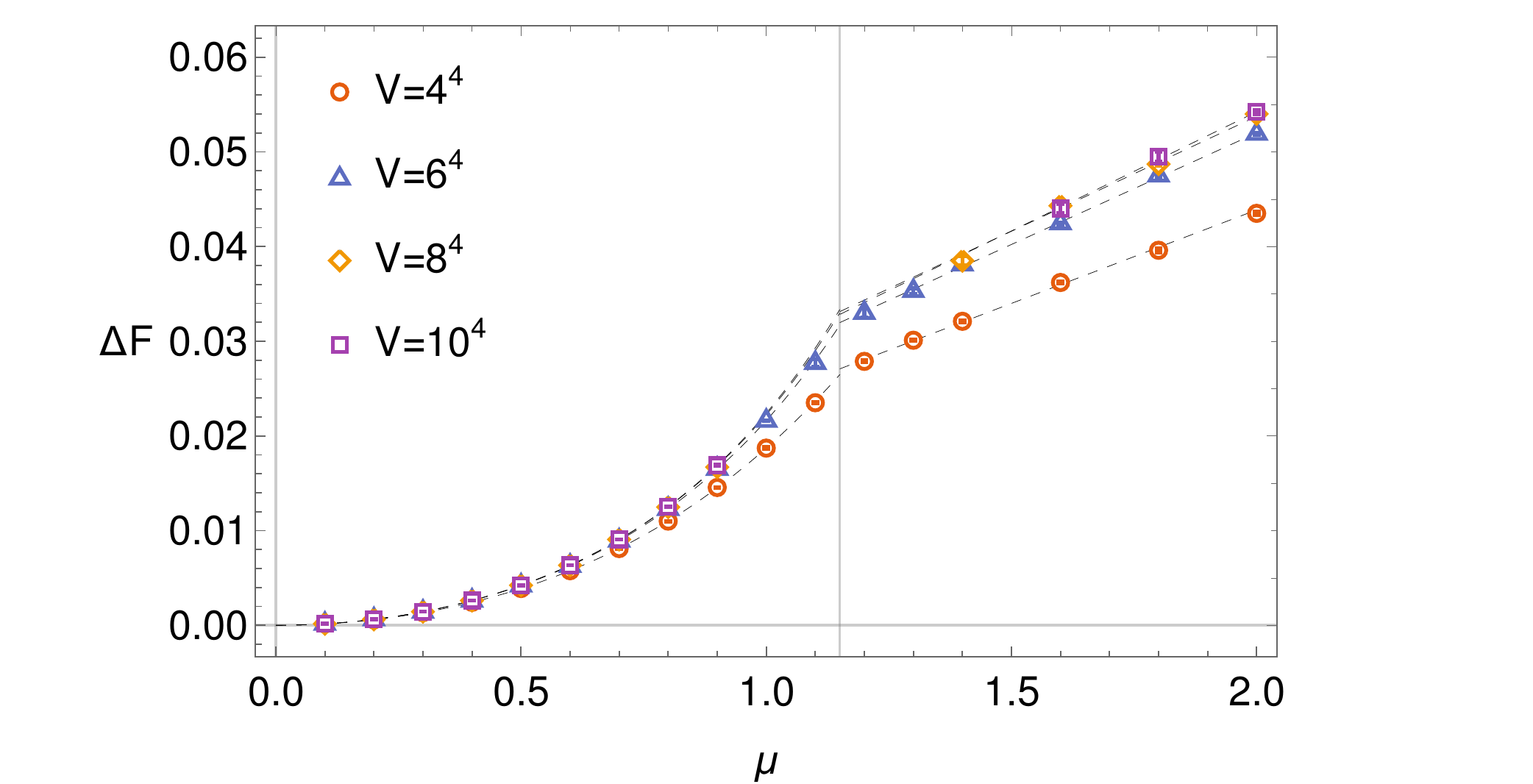}
\caption{Values of the free energy difference, obtained through the integration of the phase factor, as a function of the chemical potential for volumes $V=4^4,6^4,8^4$ and $\lambda=m=1.0$. The vertical line represent the critical value of the chemical potential obtained via mean-field calculations ($\mu_c \simeq 1.15$). The dashed lines are fit to the data meant to guide the eyes.}
\label{fig:df_low_vol}
\end{figure}

In Fig.~\ref{fig:df_low_vol} we show the results of our simulations up to $V = 10^4$ (reported in Tab.~\ref{tab:2}) and for chemical potential values ranging from zero to $\mu=2.0$ and $\lambda=m=1.0$. We stress that our procedure fails to converge for large volumes in a window close to the critical $\mu$, while no issues are found for values of $\mu$ sufficiently far from the critical $\mu$.

In the region $\mu_c \simeq 1.15$ (as predicted by mean-field analysis) we expect, and observe, a phase transition. A clear difference in the behaviour of the free energy is visible in the two phases, distinctly for $4^4$ and $6^4$, and reasonably clearly also for $8^4$ and $10^4$.

\begin{table}[htb]
\centering
\begin{tabular}{ccccc}
\multicolumn{5}{c}{$\Delta F \times 10^3$}                    \\ \hline\hline
\multicolumn{1}{c|}{}     & \multicolumn{4}{c}{Volume}        \\
\multicolumn{1}{c|}{$\mu$}& $4^4$     & $6^4$     & $8^4$     & $10^4$     \\ \cline{1-5} 
\multicolumn{1}{c|}{0.1}  & 0.1448(1) & 0.1541(1) & 0.1547(1) & 0.15473(5) \\
\multicolumn{1}{c|}{0.2}  & 0.5840(4) & 0.6233(3) & 0.6255(2) & 0.62575(9) \\
\multicolumn{1}{c|}{0.3}  & 1.337(1)  & 1.428(1)  & 1.433(1)  & 1.4344(3)  \\
\multicolumn{1}{c|}{0.4}  & 2.423(2)  & 2.599(2)  & 2.616(2)  & 2.6167(5)  \\
\multicolumn{1}{c|}{0.5}  & 3.883(3)  & 4.194(4)  & 4.225(6)  & 4.227(1)   \\
\multicolumn{1}{c|}{0.6}  & 5.761(4)  & 6.277(1)  & 6.333(2)  & 6.340(1)   \\
\multicolumn{1}{c|}{0.7}  & 8.10(2)   & 8.938(5)  & 9.06(1)   & 9.068(3)   \\
\multicolumn{1}{c|}{0.8}  & 11.00(2)  & 12.28(1)  & 12.48(1)  & 12.523(3)  \\
\multicolumn{1}{c|}{0.9}  & 14.55(4)  & 16.52(2)  & 16.82(2)  & 16.90(4)   \\
\multicolumn{1}{c|}{1.0}  & 18.7(1)   & 21.59(2)  & - -       & - -        \\
\multicolumn{1}{c|}{1.1}  & 23.50(8)  & 27.7(4)   & - -       & - -        \\
\multicolumn{1}{c|}{1.2}  & 27.87(9)  & 33.0(5)   & - -       & - -        \\
\multicolumn{1}{c|}{1.3}  & 30.10(8)  & 35.3(2)   & - -       & - -        \\
\multicolumn{1}{c|}{1.4}  & 32.1(1)   & 38.1(2)   & 38.5(5)   & - -        \\
\multicolumn{1}{c|}{1.6}  & 36.2(1)   & 42.45(9)  & 44.3(2)   & 44.0(7)    \\
\multicolumn{1}{c|}{1.8}  & 39.6(2)   & 47.5(1)   & 48.9(4)   & 49.5(7)    \\
\multicolumn{1}{c|}{2.0}  & 43.5(2)   & 51.9(1)   & 54.0(5)   & 54.2(2)    \\
\end{tabular}
\caption{Free energy difference for volumes $V= 4^4, \ 6^4, \ 8^4$ and $10^4$ in a wide range of values for the chemical potential. The $--$ identify values of the parameters for which our interpolation method did not produce a robust result.\label{tab:2}}
\end{table}

\begin{figure}[htb]
\centering
\includegraphics[width=0.7\textwidth]{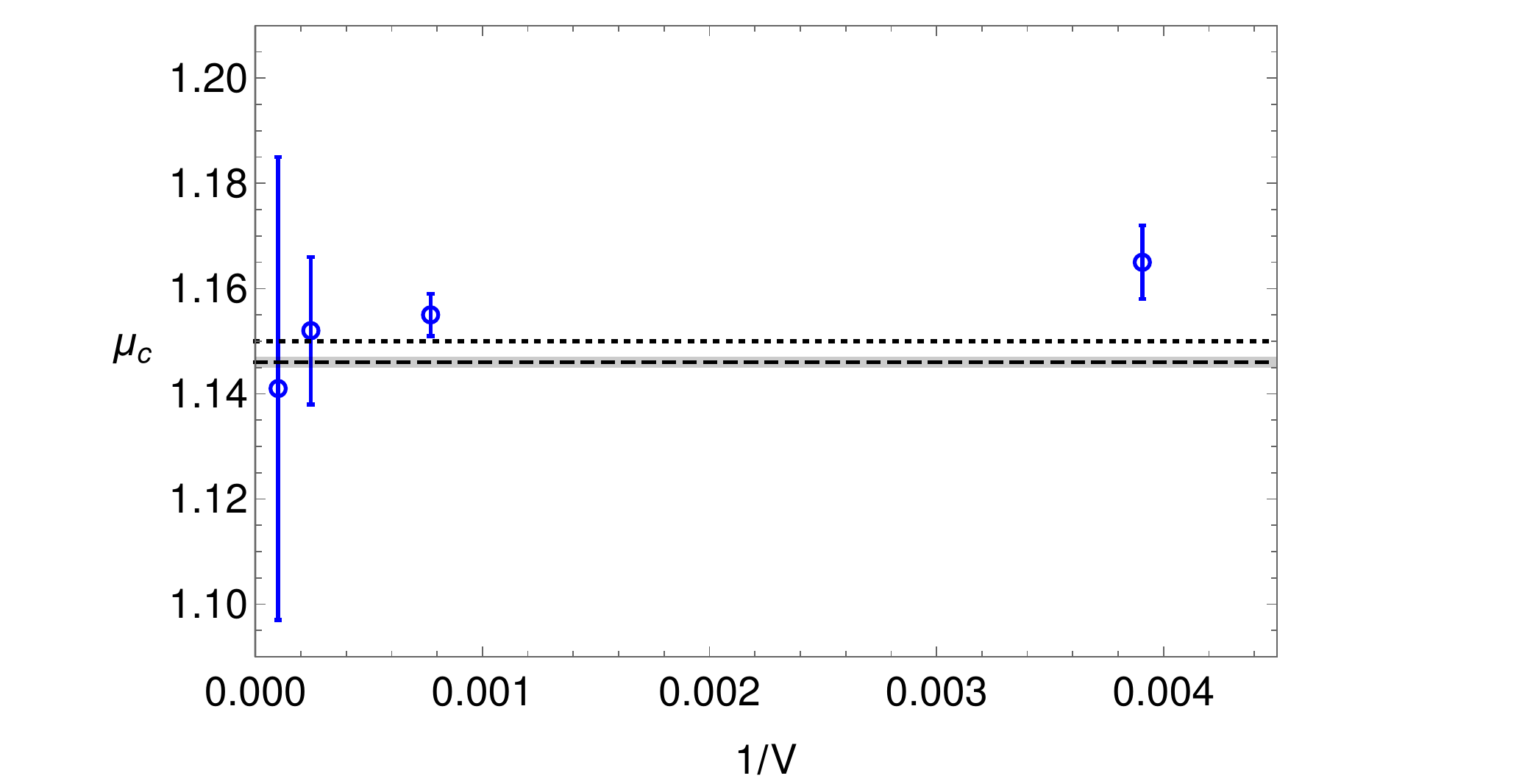}
\caption{Critical chemical potential estimates plotted against $1/V$ for volumes $V= 4^4, \ 6^4, \ 8^4$ and $10^4$. The dotted line indicates the mean-field calculation, while the dashed line is the value obtained in~\cite{Gattringer:2012df} (the error band is also indicated, but it is barely visible on the scale of the plot).}
\label{fig:mu_c}
\end{figure}

In the region $\mu<\mu_c$ the free energy difference has been fitted to the functional form $\Delta F (\mu) = a \mu^2 + b \mu^4 + c \mu^6$ , while in the $\mu>\mu_c$ a linear fit is enough to describe the behaviour of the data. By intersecting the fits in the two regions we have been able to give an estimate for the critical value of the chemical potential as well as its error via the confidence intervals of the fits. Our data are reported in Tab.~\ref{tab:3}. As shown in Fig.~\ref{fig:mu_c}, for volumes $4^4$ and $6^4$ the extrapolation obtained has a good level of precision (respectively $.6\%$ and $.4\%$ relative error), while the results for the larger volumes suffer from the lack of points close to the phase transition, resulting in relative errors of $1\%$ for $8^4$ and $4\%$ for $10^4$.

\begin{table}[htb]
\centering
\begin{tabular}{c|ccc}
$V$      & $\mu_c$ & Error & \\ \cline{1-4} 
$4^4$    & 1.165   & 0.007 &  \\
$6^4$    & 1.155   & 0.004 &  \\
$8^4$    & 1.152   & 0.014 &  \\
$10^4$   & 1.141   & 0.044 &  
\end{tabular}
\caption{Results for $\mu_c$ as a function of $V$. \label{tab:3}}
\end{table}

Since our ability to study values near $\mu_c$ decreases as the volume increases, we have not performed an extrapolation to the thermodynamic limit. However, our calculation shows that our results are compatible with the mean-field calculations \cite{Aarts:2009hn} ($\mu_c \simeq 1.15$) as well as with the value obtained in \cite{Gattringer:2012df} ($\mu_c = 1.146 \pm 0.001$) with a dual formulation of the same theory in a work more focused to the study of the phase transition than the present one. The statistical uncertainty in our result is of the order of a few percent. A careful determination of the systematic error would require an improvement of our method in order for us to be able to simulate closer to $\mu_c$ on larger lattices. 

\subsection*{Low density region}

Far from the phase transition the integration procedure poses no threat. Hence, we could study more precisely the low chemical potential region ($\mu=0\sim0.9$), extending the results to higher volumes where the sign problem get exponentially harder. The minimum polynomial order required to describe the $a_k$ data in this region ranges from 5 to 9, and for all the volumes and values of the chemical potential at least three subsequent polynomial orders (i.e. $\{7,9,11\}$) managed to integrate to statistically comparable values.

\begin{figure}[htb]
\centering
\includegraphics[width=0.7\textwidth]{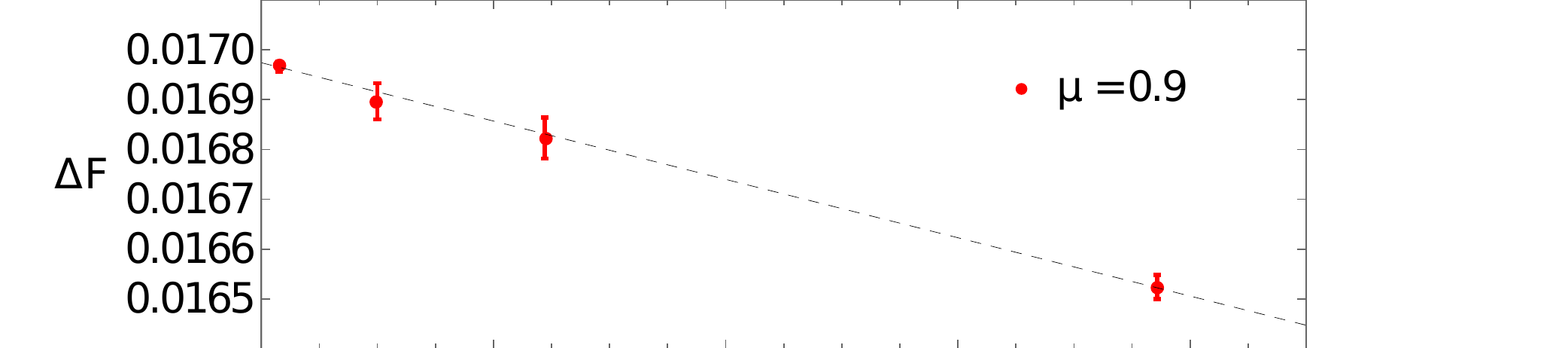}
\includegraphics[width=0.7\textwidth]{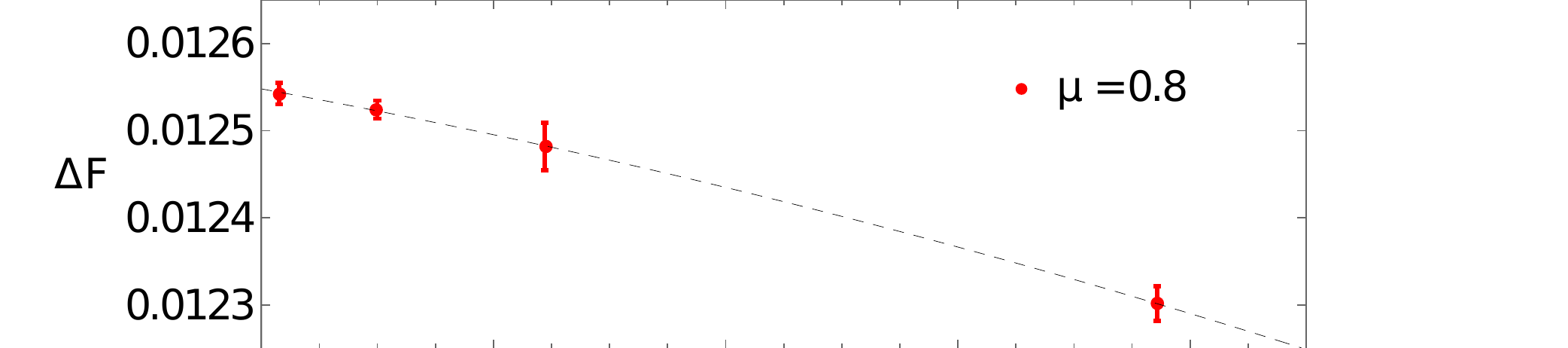}
\includegraphics[width=0.7\textwidth]{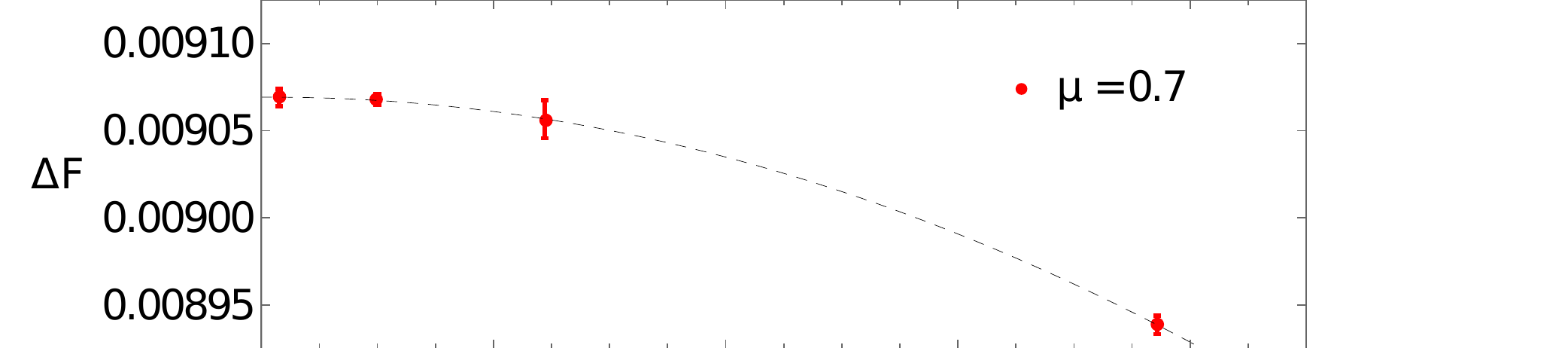}
\includegraphics[width=0.7\textwidth]{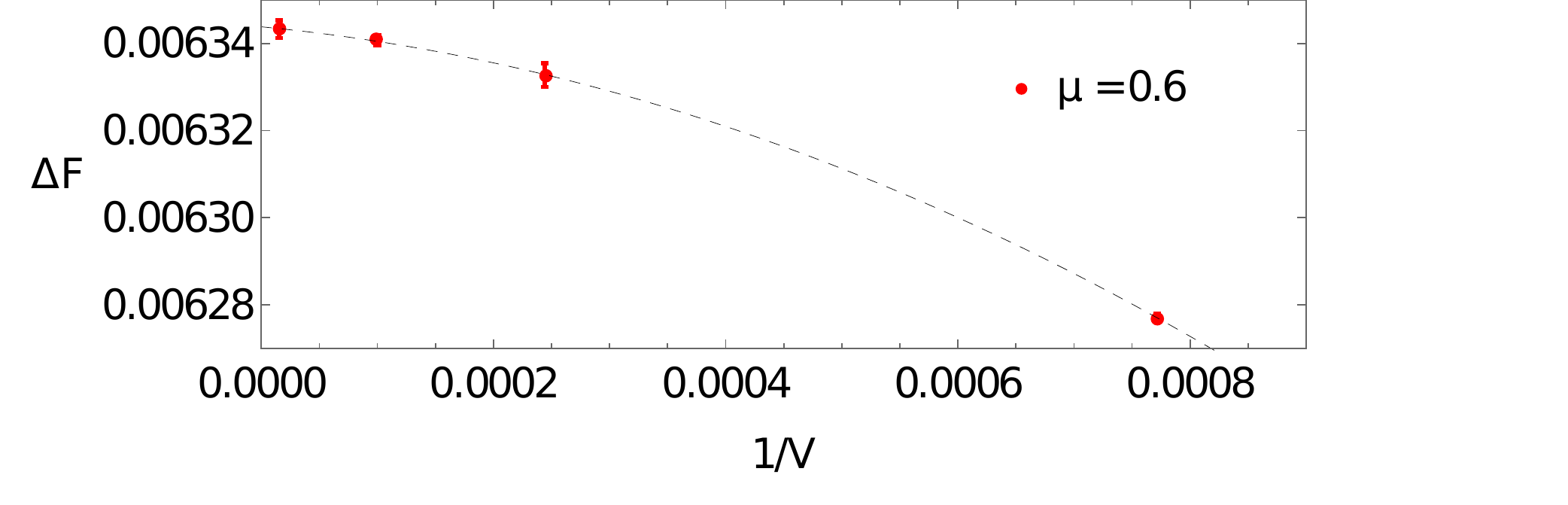}
\caption{Infinite volume scaling analysis for three values of the chemical potential $\mu=0.6,0.8,0.9$ and for volumes $V=6^4,8^4,10^4,16^4$. The fit used to extrapolate the infinite volume results are shown as well (dashed lines).\label{fig:fit_deltaf_vinfty}}
\end{figure}

\begin{figure}[htb]
\centering
\includegraphics[width=0.7\textwidth]{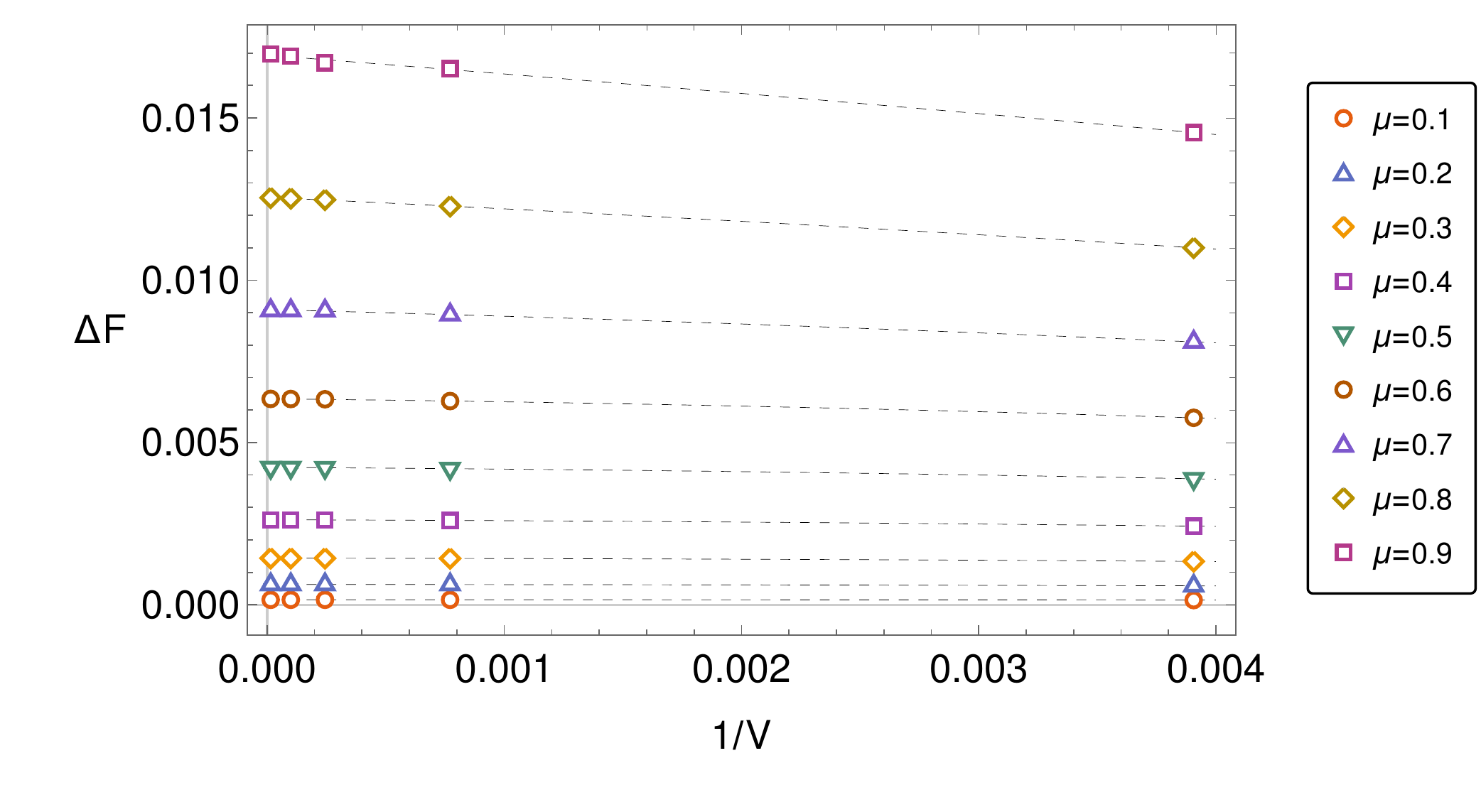}
\includegraphics[width=0.7\textwidth]{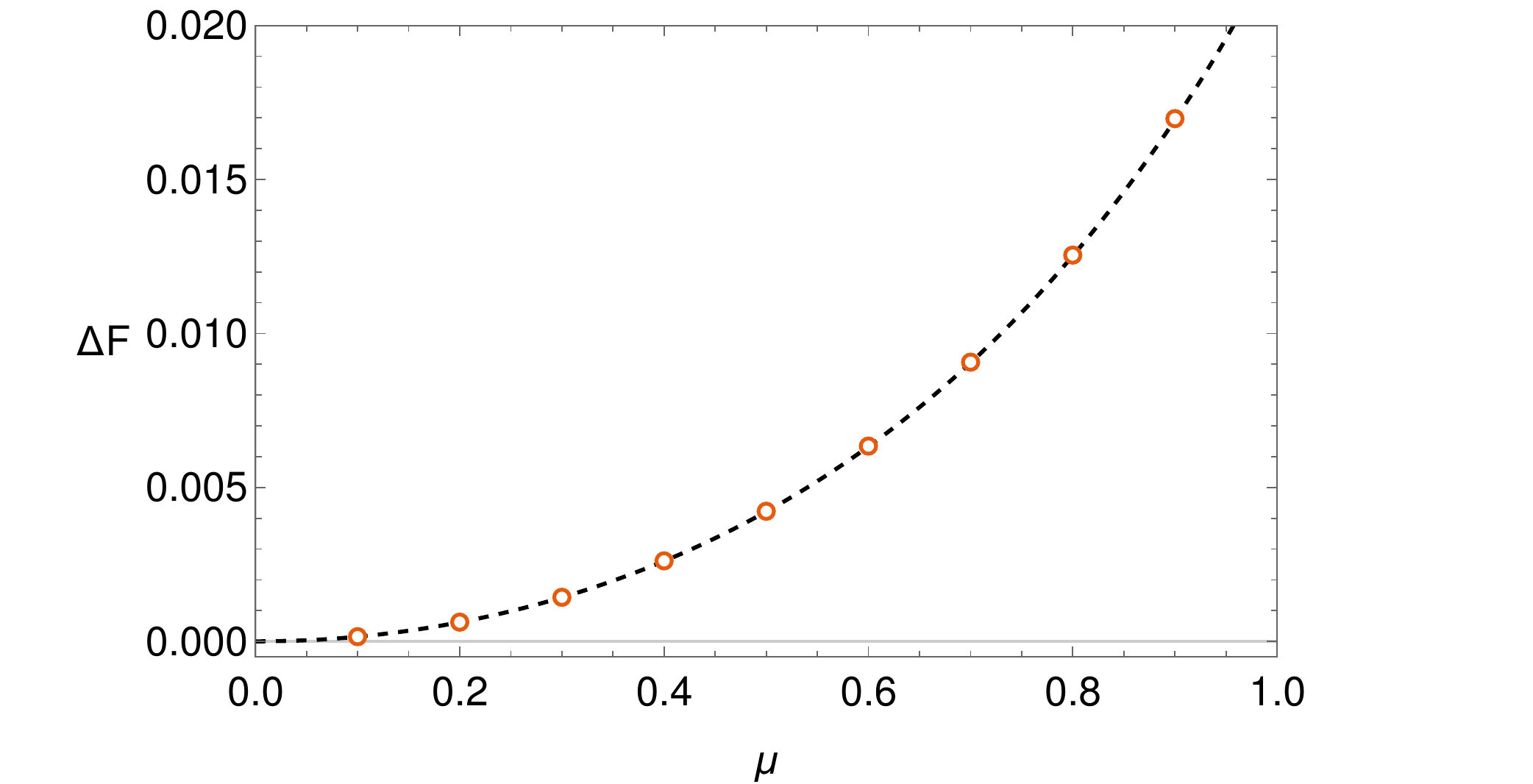}
\caption{\textbf{Top}: Free energy difference values obtained for volumes $V=4^4,6^4,8^4,10^4,16^4$, with $\lambda=m=1.0$ at values of the chemical potential in the range $\mu=\{0.1 \sim 0.9\}$; also shown is their extrapolation to the infinite volume limit. \textbf{Bottom}: Comparison between the infinite volume extrapolation obtained with the data on the top plot and the same analysis made with mean-field calculations.\label{fig:fig_deltafinfty}}
\end{figure}

\begin{table}[htb]
\centering
\begin{tabular}{ccccccc}
\multicolumn{7}{c}{$\Delta F \times 10^3$}                                                                                                                  \\ \hline\hline
\multicolumn{1}{c|}{}    & \multicolumn{5}{c|}{Volume}                                                      &                \\
\multicolumn{1}{c|}{$\mu$}& $4^4$    & $6^4$     & $8^4$     & $10^4$     & \multicolumn{1}{c|}{$16^4$}     & $\infty$ extr. \\ \cline{1-7} 
\multicolumn{1}{c|}{0.1} & 0.1448(1) & 0.1541(1) & 0.1547(1) & 0.15473(5) & \multicolumn{1}{c|}{0.15471(1)} & 0.15470(1)     \\
\multicolumn{1}{c|}{0.2} & 0.5840(4) & 0.6233(3) & 0.6255(2) & 0.62575(9) & \multicolumn{1}{c|}{0.6257(1)}  & 0.6257(1)      \\
\multicolumn{1}{c|}{0.3} & 1.337(1)  & 1.428(1)  & 1.433(1)  & 1.4344(3)  & \multicolumn{1}{c|}{1.4343(4)}  & 1.4344(4)      \\
\multicolumn{1}{c|}{0.4} & 2.423(2)  & 2.599(2)  & 2.616(2)  & 2.6167(5)  & \multicolumn{1}{c|}{2.617(1)}   & 2.617(1)       \\
\multicolumn{1}{c|}{0.5} & 3.883(3)  & 4.194(4)  & 4.225(6)  & 4.227(1)   & \multicolumn{1}{c|}{4.227(3)}   & 4.227(3)       \\
\multicolumn{1}{c|}{0.6} & 5.761(4)  & 6.277(1)  & 6.333(2)  & 6.340(1)   & \multicolumn{1}{c|}{6.343(2)}   & 6.343(2)       \\
\multicolumn{1}{c|}{0.7} & 8.10(2)   & 8.938(5)  & 9.06(1)   & 9.068(3)   & \multicolumn{1}{c|}{9.069(5)}   & 9.068(5)       \\
\multicolumn{1}{c|}{0.8} & 11.00(2)  & 12.28(1)  & 12.48(1)  & 12.523(3)  & \multicolumn{1}{c|}{12.541(6)}  & 12.546(6)      \\
\multicolumn{1}{c|}{0.9} & 14.55(4)  & 16.52(2)  & 16.82(2)  & 16.90(4)   & \multicolumn{1}{c|}{16.967(1)}  & 16.97(1)                 
\end{tabular}
\caption{Free energy difference results in the low density region ($\mu=0.1$ to $\mu=0.9$) for different volumes and infinite volume extrapolation.
\label{tab:4}}
\end{table}

We report in Tab.~\ref{tab:4} our results for $\Delta F$ as a function of $\mu$ for volumes up to $V = 16^4$. In the same table, we show also the thermodynamic extrapolation of $\Delta F$, obtained with the ansatz
\begin{equation}
\Delta F (V) = \Delta F(\infty) + \frac{a}{V} + \frac{b}{V^2} \ , 
\end{equation}
which is a good description of our data (see Fig.~\ref{fig:fit_deltaf_vinfty} for some representative examples showing the fit quality and Fig.~\ref{fig:fig_deltafinfty}, top, for a zoomed out picture of the extrapolation in the whole range of $\mu$). The behaviour of $\Delta F$ as a function of $\mu$ for $\mu < \mu_c$ is displayed in Fig.~\ref{fig:fig_deltafinfty}, bottom. 

\section{Discussion and conclusions} \label{sect:conclusions}
In this work, we have further refined the LLR method for complex action systems, studying the main sources of systematic errors in an application to the Bose gas at finite density. Using the expected scaling with the size of the imaginary action intervals for restricted sampling, we were able to eliminate for all practical purposes the error related to this discretisation. In addition, we have further investigated the necessity of interpolating the $a_k$ in order to obtain a robust result for the oscillating integral. We expect that these lessons are generalisable to other studies of complex action systems with the LLR method. Concerning the $a_k$ interpolations, we studied several possibilities. Somewhat unexpectedly, the data support the necessity of a polynomial interpolation, which has been shown to be the only one in the set of those we analysed that is able to produce a controlled result for the highly oscillating integral. In particular, we have shown that a polynomial fitting approach produces numerically stable and reliable results for phase factors down to $\order{10^{-480}}$ occurring in our model for the scenarios with the hardest sign problem we have explored. Reasons for the failure of the other studied interpolating methods have been analysed. However, at the moment it is unclear whether the polynomial interpolation would have the same degree of success on other systems. Other studies in the literature (e.g.~\cite{Langfeld:2014nta,Garron:2016noc}) also find the polynomial interpolator sufficiently accurate, although no alternative methods have been considered in these works. In our study, we also established a criterium that allows us to assess the requested order of the fitting polynomial. 

Armed with this machinery, we have then performed a numerical investigation of the free energy difference $\Delta F$ between the full and phase-quenched system. In the low density phase, we have been able to determine this observable and to extrapolate it to the thermodynamic limit for up to volume $V = 16^4$, for chemical potential values for which the sign problem is indeed hard (as already mentioned, we have successfully resolved and compared with other methods phase factors of $\order{10^{-480}}$). Our results are compatible with those obtained with Complex Langevin, mean-field calculations and dual methods. Our method also allows us to determine $\Delta F$ for $\mu$ values that put the system in the dense phase, although the method fails in a region around $\mu_c$ whose upper bound seems to increase with the volume. For the maximum volume we have simulated in the dense phase, $V = 10^4$, we have been able to extract $\Delta F$ only for $\mu \ge 1.6$ (while the critical value is $\mu_c \simeq 1.15$). The failure of the approach in the proximity of $\mu_c$  is explained by the observation that, despite the $a_k$ turn out to be very well determined, the polynomial interpolation is not able to account for a sudden change of behaviour of these coefficients as $S_I$ increases in the region that gives non-negligible contributions to the integral. We leave to future studies to understand whether a more suitable ansatz can enable us to make progress in the currently inaccessible region. Similarly, we defer to further investigations the question of whether the upper bound of the currently inaccessible region keeps increasing with $\mu$ or stabilises. 

Despite those difficulties, we have shown that the intersection of two simple interpolation ansatz for $\Delta F$ defines a critical value of $\mu$ that is accurate at the order of the percent over the range of the volumes we have simulated and compares well with the current literature. Our analysis for the determination of $\mu_c$ implicitly assumes the validity of mean-field (in fact, since we can not determine $\Delta F$ sufficiently close to $\mu_c$, we are insensitive to any potential critical behaviour beyond mean-field). In four dimensions, one expects that the critical region grows logarithmically with the volume. Hence, the critical domain possibly is very small and hidden in the region we can not access with our numerical simulations. However, it is an interesting question whether much larger lattices than those used currently in the literature (including also studies based on dual and Langevin methods) would be necessary in order to pin down the correct critical behaviour of the system. 

\section*{Acknowledgements}
We thank L. Bongiovanni, K. Langfeld and R. Pellegrini for discussions. This work has been partially supported by the ANR project ANR-15-IDEX-02. The work of BL is supported in part by the Royal Society Wolfson Research Merit Award WM170010 and by the STFC Consolidated Grant ST/P00055X/1. AR is supported by the STFC Consolidated Grant ST/P000479/1. Numerical simulations have been performed on the Swansea SUNBIRD system, provided by the Supercomputing Wales project, which is part-funded by the European Regional Development Fund (ERDF) via Welsh Government, and on the HPC facilities at the HPCC centre of the University of Plymouth.

\bibliography{bose}

\end{document}